\PassOptionsToPackage{a4paper,margin=25mm}{geometry}
\documentclass[pdflatex,sn-nature]{sn-jnl}

\usepackage{graphicx}%
\usepackage{multirow}%
\usepackage{amsmath,amssymb,amsfonts}%
\usepackage{amsthm}%
\usepackage{mathrsfs}%
\usepackage[title]{appendix}%
\usepackage{xcolor}%
\usepackage{textcomp}%
\usepackage{manyfoot}%
\usepackage{booktabs}%
\usepackage{gensymb}
\usepackage{algorithm}%
\usepackage{algorithmicx}%
\usepackage{algpseudocode}%
\usepackage{listings}%
\usepackage{pdflscape}
\usepackage{lineno}

\newcounter{extfig}

\newenvironment{extfigure}[1][]{
    \refstepcounter{extfig}%
    \begin{figure}[#1]\centering
}{\end{figure}}

\newenvironment{extfigure*}[1][]{
    \refstepcounter{extfig}%
    \begin{figure*}[#1]\centering
}{\end{figure*}}

\newcounter{exttab}

\newenvironment{exttable*}[1][]{
    \refstepcounter{exttab}%
    \begin{table*}[#1]\centering
}{\end{table*}}

\newcommand{\arcsec}{^{\prime\prime}}
\renewcommand{\tablename}{Extended Data Table}

\begin{document}

\title[Compact dusty starbursts at cosmic noon linked to high-energy neutrinos]{Compact dusty starbursts at cosmic noon linked to high-energy neutrinos}

\author*[1,2]{\fnm{Yuji} \sur{Urata}}\email{yjurata@gmail.com}
\author*[3]{\fnm{Kuiyun} \sur{Huang}} \email{kuiyun@gmail.com}
\author[4,5,6]{\fnm{Bunyo} \sur{Hatsukade}}
\author[7]{\fnm{Mansi} \sur{Kasliwal}}
\author[8,9]{\fnm{Shigeo, S.} \sur{Kimura}}
\author[4,10]{\fnm{Yuichi} \sur{Matsuda}}
\author[11]{\fnm{Yusuke} \sur{Miyamoto}}
\author[4,5]{\fnm{Hiroshi} \sur{Nagai}}
\author[4,5]{\fnm{Kouichiro} \sur{Nakanishi}}
\author[12,13,14]{\fnm{Robert} \sur{Stein}}
\affil[1]{\orgname{MITOS Science Co., LTD}, \postcode{235}, \city{New Taipei}, \country{Taiwan}}

\affil[2]{\orgdiv{Institute of Astronomy}, \orgname{National Central University}, \orgaddress{\city{Chung-Li}, \postcode{32054}, \country{Taiwan}}}

\affil[3]{\orgdiv{Center for General Education}, \orgname{Chung Yuan Christian University}, \orgaddress{\city{Taoyuan}, \postcode{32023}, \country{Taiwan}}}

\affil[4]{\orgname{National Astronomical Observatory of Japan}, \orgaddress{\city{2-21-1 Osawa, Mitaka, Tokyo}, \postcode{181-8588}, \country{Japan}}}

\affil[5]{\orgname{The Graduate University for Advanced Studies, SOKENDAI}, \orgaddress{\city{2-21-1 Osawa, Mitaka, Tokyo}, \postcode{181-8588}, \country{Japan}}}

\affil[6]{\orgdiv{Department of Astronomy, Graduate School of Science}, \orgname{The University of Tokyo}, \orgaddress{\city{7-3-1 Hongo, Bunkyo-ku, Tokyo}, \postcode{133-0033}, \country{Japan}}}

\affil[7]{\orgdiv{Cahill Center for Astrophysics}, \orgname{California Institute of Technology} \orgaddress{MC 249-17, 1216 E California Boulevard, Pasadena, CA}, \postcode{91125}, \country{USA}}

\affil[8]{\orgdiv{Frontier Research Institute for Interdisciplinary Sciences}, \orgname{Tohoku University}, \orgaddress{\city{Sendai}, \postcode{980-8578}, \country{Japan}}}

\affil[9]{\orgdiv{Astronomical Institute}, \orgname{Tohoku University}, \orgaddress{\city{Sendai}, \postcode{980-8578}, \country{Japan}}}

\affil[10]{\orgname{Ministry of Education, Culture, Sports, Science and Technology}, \orgaddress{\city{3-2-2 Kasumigaseki, Chiyoda-ku, Tokyo}, \postcode{100-8959}, \country{Japan}}}

\affil[11]{\orgdiv{Department of Electrical, Electronic and Computer Engineering}, \orgname{Fukui University of Technology}, \orgaddress{\city{3-6-1, Gakuen, Fukui}, \postcode{910-8505}, \country{Japan}}}

\affil[12]{\orgdiv{Department of Astronomy}, \orgname{University of Maryland}, \orgaddress{College Park, MD}, \postcode{20742}, \country{USA}}

\affil[13]{\orgdiv{Joint Space-Science Institute}, \orgname{University of Maryland}, \orgaddress{College Park, MD}, \postcode{20742}, \country{USA}}

\affil[14]{\orgdiv{Astrophysics Science Division}, \orgname{NASA Goddard Space Flight Center} \orgaddress{Mail Code 661, Greenbelt, MD}, \postcode{20771}, \country{USA}}

\abstract{
The origin of high-energy astrophysical neutrinos remains unresolved, and secure electromagnetic counterparts to individual events are rare despite rapid follow-up. Dusty star-forming galaxies (DSFGs) at cosmic noon ($z\sim1$--$4$) are natural cosmic-ray calorimeters, yet observational links between DSFGs and neutrinos have remained elusive. Here we report a compact-core DSFG within an IceCube localization, JCMT0402$-$0424, a quadruply lensed galaxy at $z = 2.988$ located inside the 90\% containment region of the IceCube event IC\,210922A. 
ALMA imaging and lens modeling resolve a highly magnified, compact starburst with no bright $\gamma$-ray or X-ray counterpart above current sensitivity limits.
Considering the positional agreement, the low chance-coincidence probability ($\lesssim1\%$) for such an extreme submillimeter source, the absence of equally plausible alternatives in the field, and the compact, gas-rich core revealed by ALMA, JCMT0402$-$0424 is the most plausible electromagnetic counterpart candidate within the IC\,210922A localization. In a population context, compact-core starbursts at cosmic noon can provide a non-negligible population-level contribution to the diffuse high-energy neutrino background, even though the neutrino yield from any single DSFG is modest.
This result connects high-energy neutrino production to the peak epoch of cosmic star formation, opening a new avenue to probe galaxy evolution and cosmic-ray acceleration across cosmic time.
}


\maketitle
\section{Introduction}

High-energy astrophysical neutrinos provide a unique probe of the most extreme particle accelerators in the Universe. 
Despite systematic all-sky searches with the Fermi Large Area Telescope (LAT) and extensive multiwavelength follow-ups, electromagnetic counterparts to individual high-energy neutrino events are often ambiguous.
Beyond alert-driven searches, a statistical, population-level approach constrains the dominant emitters via their integrated contribution to the diffuse flux measured by IceCube \citep{Aartsen2013Sci,Aartsen2020PRL}.
Growing observational evidence points to nearby, non-jetted active galactic nuclei (AGNs)—in particular Seyfert galaxies—as promising neutrino emitters: IceCube has reported a statistically significant excess of TeV neutrinos from NGC~1068, establishing the first compelling steady extragalactic neutrino source of this class \citep{ngc1068}. 
Follow-up population studies further suggest that X-ray--bright, non-jetted AGN subsets may contribute to the diffuse flux, while class-wide significance is still under active investigation \citep[e.g.][]{mojave,Li2022}.

In addition to steady, non-jetted AGNs, a range of transient channels—appearing in both jetted and non-jetted variants—has been proposed as potential neutrino factories, notably tidal disruption events (TDEs) and fast blue optical transients (FBOTs) \citep[e.g.][]{Stein2021,Winter2021,2019ApJ...878L..25H}. While individual cases are intriguing, population-level analyses have not yet established these classes as major contributors to the diffuse neutrino background.
By contrast, classical jet-dominated populations—$\gamma$-ray--bright blazars and prompt gamma-ray bursts (GRBs)—are constrained to be subdominant contributors to the diffuse neutrino background: stacking and cross-correlation analyses limit the blazar contribution to at most a subdominant fraction, while GRB searches place stringent upper bounds on their integrated role \citep[e.g.][]{Li2022,GRB221009A}.

Dusty star-forming galaxies (DSFGs) at cosmic noon ($z\!\sim\!1$--4) remain compelling on physical grounds. 
Their compact, gas-rich starburst cores can approach the calorimetric limit for cosmic rays, enabling efficient $pp$ interactions and neutrino production \citep{Thompson2007,Lacki2011,Tamborra2014}. 
In such systems, the neutrino output depends not only on the cosmic-ray injection power but also on confinement and transport within the dense starburst environment. 
Nevertheless, whether DSFGs can supply a major share remains debated, because in $\gamma$-ray--transparent $pp$ sources the accompanying $\pi^0$-decay emission can overshoot the extragalactic $\gamma$-ray background (EGB) measured by Fermi-LAT \citep{Murase2016,Bechtol2017}.

\section{IC\,210922A and multi-messenger follow-ups}\label{sec2}

On 2021 September 22 at 18:17:20.948~UT, the IceCube Neutrino Observatory detected a track-like high-energy neutrino event, IC\,210922A ($E_{\nu} \sim 750$~TeV) \citep{GCN30862}. This event was identified through the \texttt{ICECUBE\_Astrotrack\_GOLD} alert stream, with an estimated false-alarm rate of $0.1472~\mathrm{yr^{-1}}$ and a signalness exceeding 90\%, making it one of the most significant neutrino alerts issued in 2021. Initial automated reconstruction was subsequently refined by offline analyses, yielding an improved source localization.
The best-fit equatorial coordinates are
$\alpha_{\mathrm{J2000}} = 60.73^{+0.96}_{-0.66}$~deg and 
$\delta_{\mathrm{J2000}} = -4.18^{+0.42}_{-0.55}$~deg 
(90\% containment) \citep{GCN30862}. 

A series of rapid follow-up observations were performed across the electromagnetic spectrum in response to IC\,210922A. 
Fermi-LAT reported no catalogued $\gamma$-ray source within the 90\% localization region and found no evidence for new emission on timescales from days to years \citep{GCN30862,GCN30867}. 
Similarly, Fermi-GBM did not detect any impulsive emission consistent with a GRB-like transient around the neutrino detection time \citep{GCN30871}. 
The ground-based $\gamma$-ray observatory HAWC also reported no significant detection of steady or transient emission from the region \citep{GCN30876}. 
In the neutrino band, the ANTARES detector carried out a targeted search but found no coincident neutrino candidates within $\pm$1\,h (and extended $\pm$1\,d) window around the event \citep{GCN30875}. 
X-ray follow-up with the Neil Gehrels \textit{Swift} Observatory covered $\sim75\%$ of the 90\% localization region with a tiled 3.5\,ks exposure, yet no counterpart was identified \citep{GCN30877}. 
In addition, long-term hard X-ray monitoring data from \textit{Swift}/BAT, spanning more than 5000 days prior to the event, reveal no persistent or flaring source at the IceCube position (\ref{fig:swift_bat_monitor}). 
At optical wavelengths, the Zwicky Transient Facility (ZTF) imaged 79\% of the localization region down to a depth of $\sim21$\,mag but detected no new transient source \citep{GCN30870}. 
Spectroscopic follow-up with the Dark Energy Spectroscopic Instrument (DESI) obtained spectra for 249 galaxies within the 90\% region, corresponding to $\sim$20\% of the Bright Galaxy Sample in the field, but found no evidence for supernovae or tidal-disruption events \citep{GCN30923}.

\section{Discovery and characterization of an extremely bright submillimeter counterpart}\label{sec3}

\begin{landscape}
\begin{figure*}[t]
\centering
\includegraphics[width=1.6\textwidth]{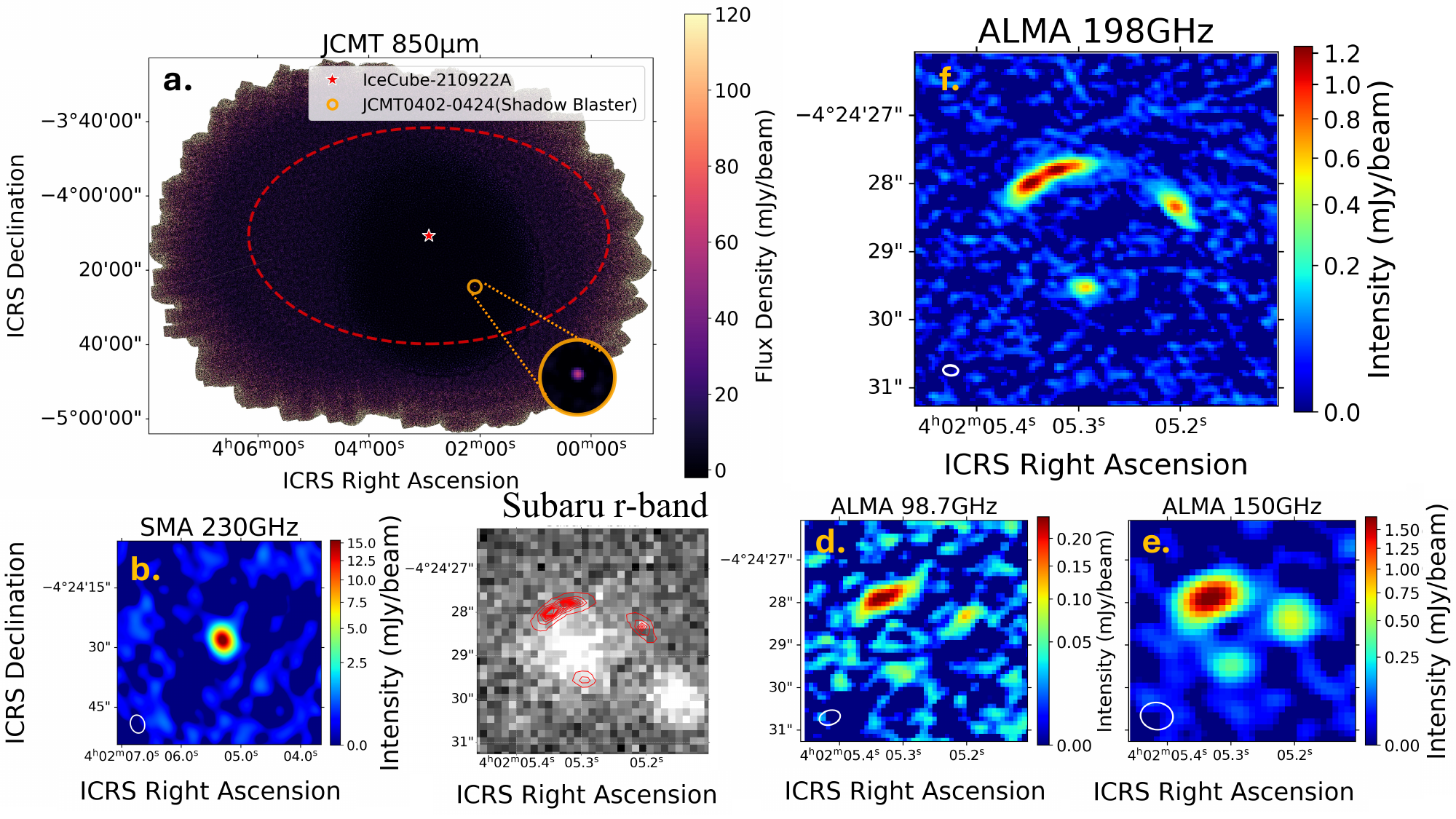}
\caption{
\textbf{Multi-wavelength identification of the IC\,210922A counterpart.} 
\textbf{a,} JCMT/SCUBA-2 850\,$\mu$m map of the IC\,210922A field. The red dashed ellipse marks the 90\% IceCube localization, and the orange circle indicates the bright submillimeter source JCMT0402$-$0424 (“Shadow Blaster”; 63\,mJy). Inset: zoom-in on the source. 
\textbf{b,} SMA 230\,GHz image providing a refined position. 
\textbf{c,} Subaru/HSC $r$-band image with ALMA Band~5 198\,GHz continuum contours overlaid, showing the foreground lensing galaxy. The contours are shown at approximately 3--11$\sigma$. 
\textbf{d,} ALMA 98.7\,GHz continuum map. 
\textbf{e,} ALMA 150\,GHz continuum map. 
\textbf{f,} ALMA 198\,GHz continuum map. The ALMA data resolve the source into four lensed images forming a quadruply imaged configuration, in contrast to the single compact appearance in the SMA map. White ellipses denote the synthesized beams. 
}
\label{fig:fig1}
\end{figure*}
\end{landscape}

Follow-up observations of the IC\,210922A localization were initiated with the James Clerk Maxwell Telescope (JCMT) using the SCUBA-2 instrument on 2021 September 24 \citep{GCN30882}. An exceptionally bright submillimeter source, designated JCMT0402$-$0424 (\textit{Shadow Blaster}; orange circle in Figure~\ref{fig:fig1}a), was identified within the 90\% point-spread-function containment region of IC\,210922A (Figure~\ref{fig:fig1}a, red dashed ellipse), with flux densities of $63 \pm 4$\,mJy at $850\,\mu \mathrm{m}$ and $168 \pm 53$\,mJy at $450\,\mathrm{\mu m}$. The SCUBA-2 follow-up observations covered the entire IceCube error region over eight epochs (September 24, 27, 30; October 6, 28; November 22, 26, 29), but no additional comparably bright sources were detected and no significant flux variability was observed. 
The probability of finding a submillimeter source brighter than 50 mJy by chance within the IC\,210922A localization is only $\sim$1\%, based on number counts from nearly all available JCMT/SCUBA-2 observations \citep{francesco}. Recent blank-sky surveys \citep{geach,garratt} indicate an even lower surface density of bright ($>50$ mJy) 850-$\mu$m sources, of order $10^{-2}$ deg$^{-2}$, implying that the true chance-coincidence probability for such a source to fall within the IceCube localization is closer to $\sim$0.3\%.
These probabilities are derived from blank-field number counts and thus characterize the expected rate of such bright submillimeter galaxies in any random region of this size. The presence of JCMT0402$-$0424 within the IC\,210922A localization is therefore statistically uncommon ($p \simeq 3\times10^{-3}$), although a chance alignment cannot be strictly excluded.
These considerations make JCMT0402$-$0424 the most plausible electromagnetic counterpart candidate within the IC\,210922A localization.

The relatively large SCUBA-2 beam size ($14''$ FWHM) limited the ability to identify multi-wavelength counterparts. Subsequent Submillimeter Array (SMA) observations refined the position to sub-arcsecond accuracy ($\sim$0.5'') \citep{GCN30891}, enabling robust associations with catalogued sources: the infrared source WISEA\,J040205.29$-$042428.5 and the radio source NVSS\,J040205$-$042434. Archival Subaru/Hyper Suprime-Cam (HSC) imaging further revealed a faint optical counterpart coincident with the SMA position (Figure~\ref{fig:fig1}c).
Despite the refined source localization, pointed follow-up observations with \textit{Swift}/XRT and \textit{NuSTAR} did not reveal any associated X-ray emission, placing upper limits: for \textit{Swift}/XRT, the $3\sigma$ upper limit on the 0.3--10~keV flux is $8.5\times10^{-14}\,\mathrm{erg\,cm^{-2}\,s^{-1}}$ (absorbed; photon index $\Gamma=2$, see Methods), while for \textit{NuSTAR} the $3\sigma$ upper limit on the flux in the 3--70~keV band is $1.6\times10^{-11}\,\mathrm{erg\,cm^{-2}\,s^{-1}}$ (photon index $\Gamma=2$) \citep{GCN30937}.
Additional observations with VERA yielded no detection, with a $3\sigma$ upper limit of $\sim$100 mJy at 22 GHz.


High-resolution observations with ALMA in Band~3 (98.7 GHz), Band~4 (150 GHz), and Band~5 (198 GHz) revealed that JCMT0402$-$0424, which appeared compact in the SMA image, is in fact a strongly lensed system (Figure~\ref{fig:fig1}d--f).
The continuum imaging in Band~5 resolves the source into four distinct lensed images, 
including elongated arcs, forming a typical quadruply imaged configuration produced by strong gravitational lensing 
(Figure~\ref{fig:fig1}f). 
A comparison with the Subaru/HSC optical image shows a galaxy located near the geometric center of the lensed configuration (Figure~\ref{fig:fig1}c). 
Lens modeling of the relative positions and fluxes of the four images further confirms that the lens mass centroid coincides with this galaxy, supporting its identification as the lensing galaxy (see Methods).

In addition to the continuum, the ALMA spectra across three frequency bands clearly detected multiple molecular and atomic lines: CO(3--2) in Band~3, CO(5--4) in Band~4, and in Band~5 both mid- to high-$J$ CO lines (CO(6--5) and CO(7--6)) together with [C\,\textsc{i}](2--1) (Figure~\ref{fig:fig2}). 
Each line is detected at high signal-to-noise ratio and was individually modeled with a single-Gaussian profile to determine the line center (\ref{fig:method_redshift}). 
The observed line frequencies yield mutually consistent redshifts, providing a robust spectroscopic determination of $z = 2.9880 \pm 0.0003$, placing the source at the peak epoch of cosmic star formation activity ($z \sim 3$).
The statistical uncertainties correspond to a precision of $\sigma_z \simeq (1$--$5)\times 10^{-4}$, depending on the transition.

The broadband spectral energy distribution (SED) of the source is shown in 
Figure~\ref{fig:fig3} (left panel, blue points), where it is compared to that of the 
archetypal DSFG SMM\,J2135$-$0102 (the “Eyelash” galaxy) at $z=2.3258$ \citep[e.g.,][]{Swinbank2010}. 
The close similarity between the two SEDs demonstrates that the IC\,210922A counterpart shares the 
characteristic features of compact starburst-core DSFGs, with a steeply rising submillimeter continuum 
peaking in the rest-frame far-infrared. 
This resemblance supports the interpretation of the source as a highly magnified, compact star-forming galaxy at $z\sim3$.

\section{Gravitational lens modeling and intrinsic properties}

The quadruply imaged morphology revealed by ALMA demonstrates that the system is strongly lensed, and characterization of the lens galaxy is required to determine the intrinsic properties of JCMT0402$-$0424. 
The lensing galaxy, visible in the Subaru/HSC image, has a photometric redshift of $z_{\rm lens}=1.06\pm0.04$ derived from optical and near-infrared broadband photometry (\ref{fig:photoz_sed}).
Additional Gemini/GNIRS near-infrared spectroscopy further supports this identification, showing no emission features, consistent with an evolved elliptical galaxy at this redshift (\ref{fig:gnirs_spec}).

Gravitational lens modeling of the four-image configuration shows that the lens galaxy is accurately reproduced by an elliptical power-law mass profile with Einstein radius $\theta_E = 1.068 \pm 0.006$~arcsec, corresponding to an enclosed mass of $M(<\theta_E) = (1.1 \pm 0.1) \times 10^{11}$~M$_\odot$ at the lens redshift $z=1.06$ (\ref{fig:lens_corner}). 
The mass profile slope $\gamma = 1.97 \pm 0.11$ is consistent with an isothermal distribution typical of massive elliptical galaxies, while the modest ellipticity and external shear indicate a relatively relaxed environment.
The background source, JCMT0402$-$0424, is modeled with two components: an extended S\'{e}rsic profile with effective radius $R_{\rm eff} = 0.065 \pm 0.007$~arcsec (corresponding to an intrinsic size of $\sim$520~pc) and S\'ersic index $n = 2.90 \pm 0.18$, plus an unresolved point source offset by $\sim$50~mas from the extended emission centroid. The combined model provides a good fit with $\chi^2/{\rm dof}=1.40$ (Figure \ref{fig:phase5_paper_figure}, \ref{fig:model_comparison}). This two-component structure is characteristic of high-redshift DSFGs and consistent with detailed studies of other strongly lensed systems such as SDP.81 and SMM\,J2135$-$0102 \citep[e.g.,][]{Swinbank2010,alma2015,Falgarone2017}. 
Magnification factors of $\mu_{\rm point}=26.9\pm3.5$ for the point-source component and $\mu_{\rm ext}=12.2\pm1.5$ for the extended emission were inferred. 

The spectral energy distribution was fitted with a modified blackbody model to determine the total infrared luminosity. 
After correcting for the lensing magnification of the extended component ($\mu_{\rm ext}=12.2\pm1.5$), the intrinsic luminosity is 
$L_{\rm IR} \simeq 2.7 \times 10^{12}\,L_\odot$, 
placing the galaxy firmly in the regime of ultra-luminous infrared galaxies (ULIRGs). 
This corresponds to an intrinsic star formation rate of 
$\sim270$--470~M$_\odot$\,yr$^{-1}$ depending on the adopted initial mass function, 
consistent with vigorous starburst activity. 
The derived luminosity and compact morphology together indicate that JCMT0402$-$0424 is a highly magnified DSFG undergoing intense star formation at $z\sim3$.

These luminosity and star formation rate estimates provide a global view of the system. 
To further constrain the baryonic content and internal dynamics, the molecular gas reservoir and dynamical mass were investigated using CO and [C\,\textsc{i}] line diagnostics \citep[e.g.,][]{Solomon1997,Solomon2005,CarilliWalter2013,Papadopoulos2004}. 
Multi-Gaussian decomposition of the ALMA spectra (Figure~\ref{fig:fig2}, Extended Data Table~1) reveals a complex velocity structure, with the broad component ($\mathrm{FWHM} \sim 400$--$500\,\mathrm{km\,s^{-1}}$) providing the dominant contribution across all CO transitions. The inferred line widths are typical of compact starburst DSFGs at $z\sim3$ \citep[e.g.,][]{Bothwell2013} 
and are significantly narrower than the extreme ($\gtrsim 1000\,\mathrm{km\,s^{-1}}$) molecular outflows often associated with powerful active galactic nuclei (AGN) \citep[e.g.,][]{Sturm2011,Cicone2014}. 
Adopting a CO(3--2)/CO(1--0) brightness temperature ratio 
\(r_{31} = 0.5\) \citep[e.g.,][]{CarilliWalter2013,Bothwell2013}  and a CO-to-H\(_2\) conversion factor 
\(\alpha_{\mathrm{CO}} = 0.8\) appropriate for starburst galaxies \citep{DownesSolomon1998}, 
the total molecular gas mass is estimated as 
\(M_{\mathrm{gas}} \simeq 7 \times 10^9\,M_\odot.\)
An independent estimate from the [C \textsc{i}] line yields a somewhat larger 
value of \(M_{\rm gas}\sim(4\text{--}10)\times10^{10}\,M_\odot\), assuming an excitation temperature 
of \(T_{\rm ex}=30\) K and a [C \textsc{i}]-to-H\(_2\) abundance ratio 
\(X_{\rm CI}=3\times10^{-5}\) \citep[e.g.,][]{Weiss2003, Papadopoulos2004}, 
reflecting systematic uncertainties in abundance and excitation.
%
The dynamical mass, inferred from the CO line widths and the lensing-corrected source-plane radius of $\sim0.5$\,kpc, is $M_{\rm dyn}\sim(1$--$3)\times10^{10}\,M_\odot$. 
This value is broadly consistent with the gas-mass estimates within the uncertainties of conversion factors and lensing magnification. 
The comparison indicates that molecular gas constitutes a substantial fraction of the total mass within the central region. 
The broad CO component dominates the reservoir, 
demonstrating that the molecular gas is strongly concentrated in the compact starburst core.

Taken together, the close SED similarity to archetypal systems such as SMM\,J2135$-$0102, the measured CO line widths (FWHM $\sim$400--500\,km\,s$^{-1}$), the strong central gas concentration, and the lensing-resolved continuum morphology indicate that the submillimeter emission of JCMT0402$-$0424 is primarily powered by vigorous star formation, with no evidence for an energetically dominant AGN in the current data. 
Furthermore, the CO spectral line energy distribution (SLED) is strongest at mid-$J$ and declines for $J_{\rm upper}\gtrsim5$, consistent with compact--starburst excitation and lacking the sustained high-$J$ excess typical of AGN-dominated sources (\ref{fig:co_sled}; see Methods). No optical or X-ray counterpart is detected, which is expected in this scenario and remains fully consistent with a heavily obscured starburst; any AGN, if present, is likely subdominant or deeply buried given the current limits.

\section{Compact-core dusty star-forming galaxies as neutrino sources}

The compact, gas-rich starburst core revealed by lens modeling provides the dense target material and magnetic confinement required for efficient cosmic-ray interactions, fulfilling the calorimetric condition for neutrino production \citep[e.g.,][]{Thompson2007,Lacki2011,Tamborra2014}. 
The magnification-boosted expectation for JCMT0402$-$0424 in our fiducial compact-core model is at most an expected number of detected events $N_\nu \lesssim 6\times10^{-2}$ over ten years. This is small in absolute terms, but comparable to per-source expectations obtained when modeling other proposed neutrino associations such as the TDE AT2019dsg and flares of the blazar TXS~0506+056, for which typical models also predict $N_\nu \ll 1$ per event, often of order $10^{-2}$--$10^{-1}$ \citep[e.g.][]{Winter2021,Oikonomou2019, Murase2020}. In this sense, JCMT0402$-$0424 is consistent with other plausible high-energy neutrino sources on an object-by-object basis. However, compact dusty star-forming galaxies differ fundamentally from such rare, transient flares: they belong to a numerous parent population that peaks in abundance during cosmic noon, when the Universe was rapidly forming stars. As a result, the collective contribution of compact-core DSFGs, like JCMT0402$-$0424, is expected to provide a meaningful population-level contribution to the diffuse astrophysical neutrino flux \citep{Abbasi2025}.
Because compact DSFGs arise from a numerous parent population, the association is also naturally interpreted in a statistical sense. In such a population, rare systems---including strongly magnified sources---may preferentially populate the high observed-flux tail, in the sense of Eddington bias. From this perspective, the modest expectation value for any single source does not preclude the detection of an individual event from the bright end of the population.

Indeed, compact starburst cores are a common feature of DSFGs, with submillimeter interferometric surveys finding effective radii of order $\sim$1~kpc or less in a substantial fraction of systems \citep[e.g.,][]{Simpson2015,Hodge2016,Fujimoto2017}. 
Compact-core DSFGs are primarily starburst-driven with little AGN contamination, whereas dusty galaxies selected in the mid-infrared (DOGs) represent a distinct, AGN-dominated population \citep[e.g.,][]{Toba2015,Fukuchi2025}. It is therefore appropriate to treat these two populations separately when modeling their respective contributions to the diffuse neutrino background.

When the luminosity function of DSFGs is divided into compact-core and extended populations \citep{Casey2018,Zavala2021}, the compact-core DSFG population is found to 
provide a meaningful but subdominant contribution to
the diffuse astrophysical neutrino flux, typically at the $\sim$15\% level and reaching at most $\lesssim$20\% over the $\sim$30~TeV--PeV range in the DSFG-only models considered here (Figure~\ref{fig:dsfg_only}) \citep{Abbasi2025}.
This contribution arises from their large cosmic abundance and the sustained star-forming activity of compact DSFGs, particularly around cosmic noon, even though individual systems are intrinsically faint neutrino sources.
In this interpretation, the compact-core component does not imply that all DSFGs are equally neutrino-active, but rather represents a compact, neutrino-efficient subset within a morphologically diverse parent population, plausibly associated with merger-driven compaction or similarly extreme short-lived phases.
At the highest energies, the DSFG component alone tends to underpredict the central IceCube best-fit spectrum \citep{Abbasi2025}, whereas adding an obscured AGN population, represented by DOGs, yields a combined spectrum consistent with the observed level and slope (\ref{fig:dsfg_plus_agn}). 
These results indicate that, although the neutrino yield from any single DSFG is modest, compact starburst cores at cosmic noon can make a  meaningful but subdominant contribution to the diffuse high-energy neutrino background.
In this context, the non-detection of temporally or positionally coincident X-ray/$\gamma$-ray emission is not, by itself, evidence for a $\gamma$-ray--dark hadronic channel, but is expected for a DSFG at $z\sim3$ given current sensitivities. The viability of the hadronic scenario is instead assessed through a population-level comparison with the Fermi-LAT EGB residual (Methods).

Considering the positional likelihood and source properties---together with the absence of comparably plausible alternatives in the field---JCMT0402$-$0424 is the most plausible counterpart candidate within the IC\,210922A localization. 
%
The single-source event expectation inferred from its $L_{\rm IR}$ is modest, while the ensemble of compact-core DSFGs sustains a measurable share of the diffuse neutrino flux.
In this framework, strongly lensed DSFGs such as JCMT0402$-$0424 are the most readily identifiable representatives of this population within IceCube localizations with current observational capabilities. However, the bulk of the contribution is expected to arise from unlensed compact-core DSFGs with submillimeter flux densities of order a few mJy. Systematic identification of such sources across the wide IceCube error regions is not yet feasible, because no current facility can survey these areas to the required depth within realistic observing-time constraints.
Wide-field, mJy-level submillimeter surveys—for example with future large single-dish telescopes such as the proposed Large Submillimeter Telescope (LST)—will therefore be crucial to test the broader role of DSFGs as neutrino sources.

\section{Summary}\label{sec6}

The quadruply lensed DSFG JCMT0402$-$0424 at $z = 2.988$, located within the IceCube event IC\,210922A error region, represents a compelling compact-core DSFG proposed as a high-energy neutrino counterpart candidate. Multi-wavelength follow-up and gravitational lens modelling reveal a compact, gas-rich starburst core with $L_{\rm IR} \simeq 2.7 \times 10^{12}\,L_\odot$, a star formation rate of $270$--$470\,M_\odot\,{\rm yr^{-1}}$, and a molecular gas reservoir consistent with an extreme starburst environment. 
The compactness and high gas surface density satisfy calorimetric conditions for efficient cosmic-ray interactions, while the absence of a bright $\gamma$-ray or X-ray counterpart above current sensitivity limits is expected for a source at this distance. 
Taken together with population modelling of compact-core DSFGs and their consistency with the Fermi-LAT EGB residual constraint, these results indicate that although the neutrino yield expected from any single compact-core DSFG is modest, the population can provide a meaningful but subdominant contribution to the diffuse neutrino background, while strongly lensed DSFGs serve as natural laboratories for exploring the connection between extreme star formation and high-energy particle acceleration at cosmic noon.

\section*{Methods}
\addcontentsline{toc}{section}{Methods} 

Unless otherwise noted, all quoted uncertainties correspond to 1$\sigma$ statistical errors.

\subsection*{JCMT observations and analysis}

We observed the field of IC\,210922A with the SCUBA-2 instrument on the James Clerk Maxwell Telescope (JCMT) under program M21BP064. The first observations, obtained on 2021 September 24 in standard daisy scan mode, 
revealed a bright submillimeter source with flux densities of 
$63\pm4$\,mJy at 850\,$\mu$m and $168\pm53$\,mJy at 450\,$\mu$m. 
The source position measured in the SCUBA-2 map is 
RA(J2000) = 04:02:05.362, Dec(J2000) = $-04$:24:28.80. 
We hereafter refer to this object as JCMT0402$-$0424 (nicknamed “Shadow Blaster”), 
as labeled in Figure~\ref{fig:fig1}a.
The JCMT beam size of $\sim$14'' (FWHM at 850\,$\mu$m) limits the positional accuracy, motivating higher-resolution follow-up with the SMA and ALMA.
Subsequent monitoring over eight epochs between September and November 2021 
showed no significant variability within the statistical uncertainties, 
confirming the source as persistent rather than a transient counterpart.

The SCUBA-2 mapping covered the entire 90\% containment region of the IceCube localization (Figure~\ref{fig:fig1}a). In the central region of the maps, the co-added rms sensitivities reach $\sim$2.5\,mJy at 850\,$\mu$m and $\sim$40\,mJy at 450\,$\mu$m, with typical atmospheric opacities $\tau_{225}=0.07$-0.08, indicative of favorable submillimeter conditions. 

The data were reduced with the \textsc{Starlink/SMURF} pipeline \citep{Chapin2013}, using the iterative \texttt{makemap} procedure that includes flat-fielding, extinction correction, and removal of common-mode noise. Flux calibration was performed using nightly observations of standard JCMT calibrators (e.g. Uranus, CRL\,618) following the SCUBA-2 calibration plan \citep{Holland2013}, yielding an absolute flux accuracy of 5-10\%.

\subsection*{SMA observations and analysis}

We used the Submillimeter Array (SMA) to refine the position of JCMT0402$-$0424. 
A dual-receiver setup at 230\,GHz was executed on 2021 September 27 from 15{:}21 to 17{:}38\,UT. 
Both receivers were tuned in the 230\,GHz band with upper-sideband rest frequencies 
$230.538$\,GHz (Rx~A) and $238.538$\,GHz (Rx~B), respectively. 
Weather conditions satisfied the requested PWV$<4$\,mm. 
Eight antennas were scheduled, with one unavailable during the track due to maintenance.

Standard calibration procedures were followed: BL~Lac served as the bandpass calibrator, 
0423$-$013 as the complex-gain (phase) calibrator, and Neptune as the flux calibrator. 
The data were reduced with the SMA \textsc{MIR} pipeline and imaged in \textsc{MIRIAD} \citep{MIRIAD}. 
Flagging removed corrupted visibilities before solving for gains on 0423$-$013 and transferring solutions to the target. 

The source was clearly detected at 
RA(J2000) $=$ 04$^{\rm h}$02$^{\rm m}$05.304$^{\rm s}$, 
Dec(J2000) $=$ $-04^{\circ}$24$'$28.14$''$, 
with a peak flux density of $16.91\pm0.17$\,mJy\,beam$^{-1}$ at 230\,GHz. 
The synthesized beam was $4.6''\times3.5''$ (PA $=15^\circ$). 
Given the high detection significance (S/N $\sim$100), the astrometric accuracy is estimated to be $\sim0.5''$. 
This position is consistent with the infrared source WISEA\,J040205.29$-$042428.5 and, within the astrometric uncertainties, 
with the radio source NVSS\,J040205$-$042434. 
These interferometric coordinates provided the first robust cross identification of JCMT0402$-$0424 across the infrared and radio bands.

\subsection*{ALMA observations and analysis}

JCMT0402$-$0424 was observed with ALMA (Project 2023.1.01229.S) in Bands~3, 4, and 5 
between 2023 December and 2024 July using the 12-m array in compact configurations 
(C43-1 to C43-6). Several execution blocks did not pass QA0 and were discarded, and only the datasets that successfully passed QA2 were retained for analysis. 
These datasets were processed with the standard ALMA Pipeline 
by the ALMA Observatory and imaged using CASA \citep{casa1,casa2}.
The resulting synthesized beams and continuum sensitivities for each band were as follows: 
Band 3 — \(0.50'' \times 0.34''\) with 0.022 mJy beam\(^{-1}\); 
Band 4 — \(0.77'' \times 0.64''\) with 0.032 mJy beam\(^{-1}\); 
Band 5 — \(0.28'' \times 0.17''\) (173 GHz) and \(0.22'' \times 0.15''\) (198 GHz), 
with 0.04 mJy beam\(^{-1}\) continuum sensitivities.
High-resolution continuum imaging in Band~5 resolves JCMT0402$-$0424 into four 
distinct lensed images, forming a quadruply imaged configuration centered on a 
foreground optical galaxy (Figure \ref{fig:fig1}). 
This quadruply imaged morphology provided the basis for gravitational lens modeling (see the section Gravitational lens modeling).

\begin{landscape}
\begin{figure*}[t]
  \centering
  \includegraphics[width=1.4\textwidth]{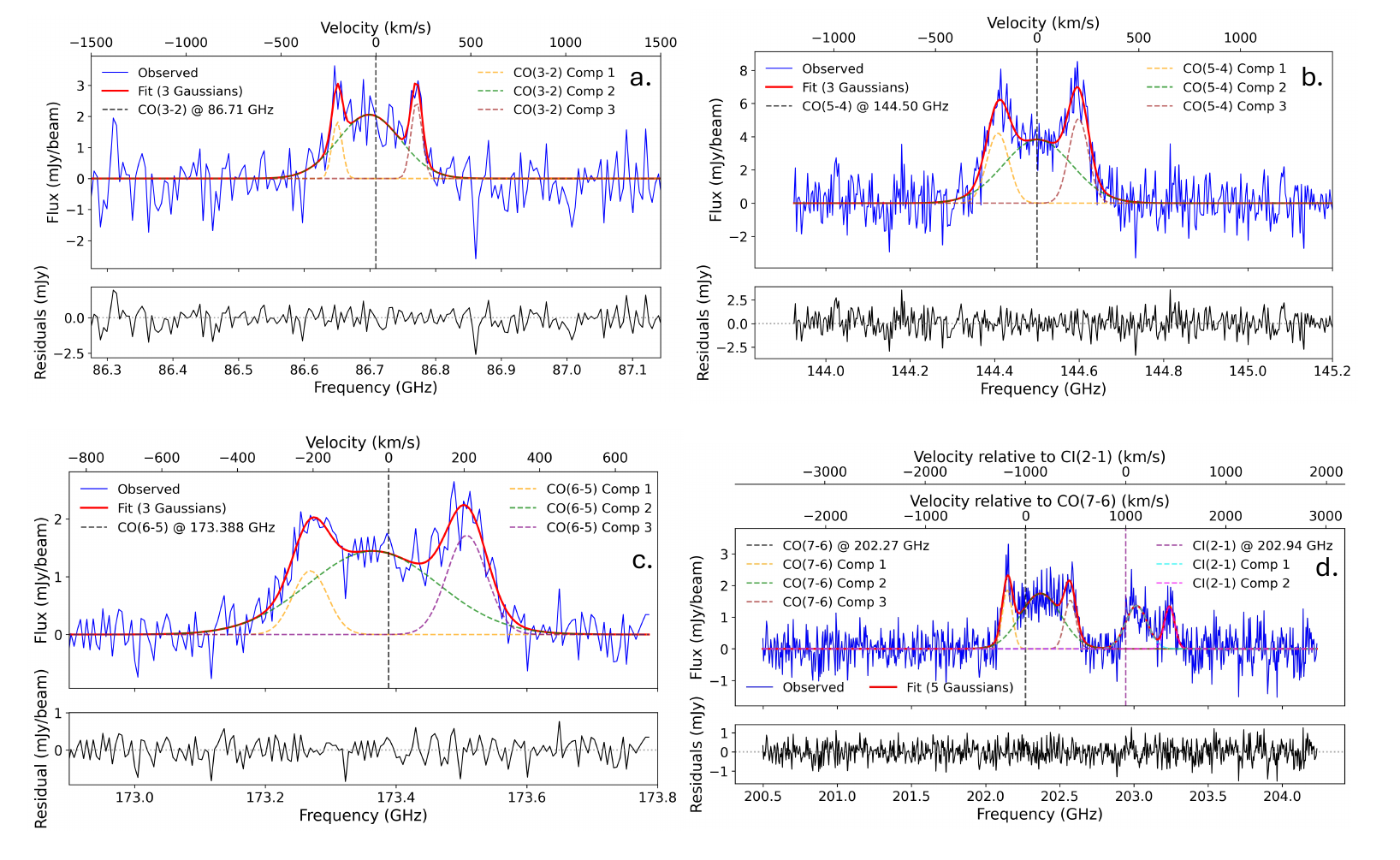}
\caption{
\textbf{ALMA spectra and multi-component Gaussian fits.}
\textbf{a,} Band~3 CO(3--2) spectrum, modeled with three Gaussian components (narrow--broad--narrow). 
\textbf{b,} Band~4 CO(5--4) spectrum, also described by three Gaussian components. 
\textbf{c,} Band~5 CO(6--5) spectrum, modeled with three Gaussian components. 
\textbf{d,} Band~5 spectrum including CO(7--6) and [C\,\textsc{i}](2--1), requiring five Gaussians (three CO components and two [C\,\textsc{i}] components). 
In each panel, the observed spectrum is shown in blue, the total fit in red, individual Gaussian components as dashed lines, and the residuals (data--model) in black below. 
Vertical dashed lines mark the expected observed frequencies (CO(3--2) 86.71\,GHz; CO(5--4) 144.50\,GHz; CO(6--5) 173.39\,GHz; CO(7--6) 202.27\,GHz; [C\,\textsc{i}](2--1) 202.94\,GHz). 
Fitting results, including $\chi^2/\mathrm{dof}$, are summarized in Extended Data Table~1.
}
  \label{fig:fig2}
\end{figure*}
\end{landscape}

\begin{landscape}
\begin{figure*}[t]
  \centering
  \includegraphics[width=1.6\textwidth]{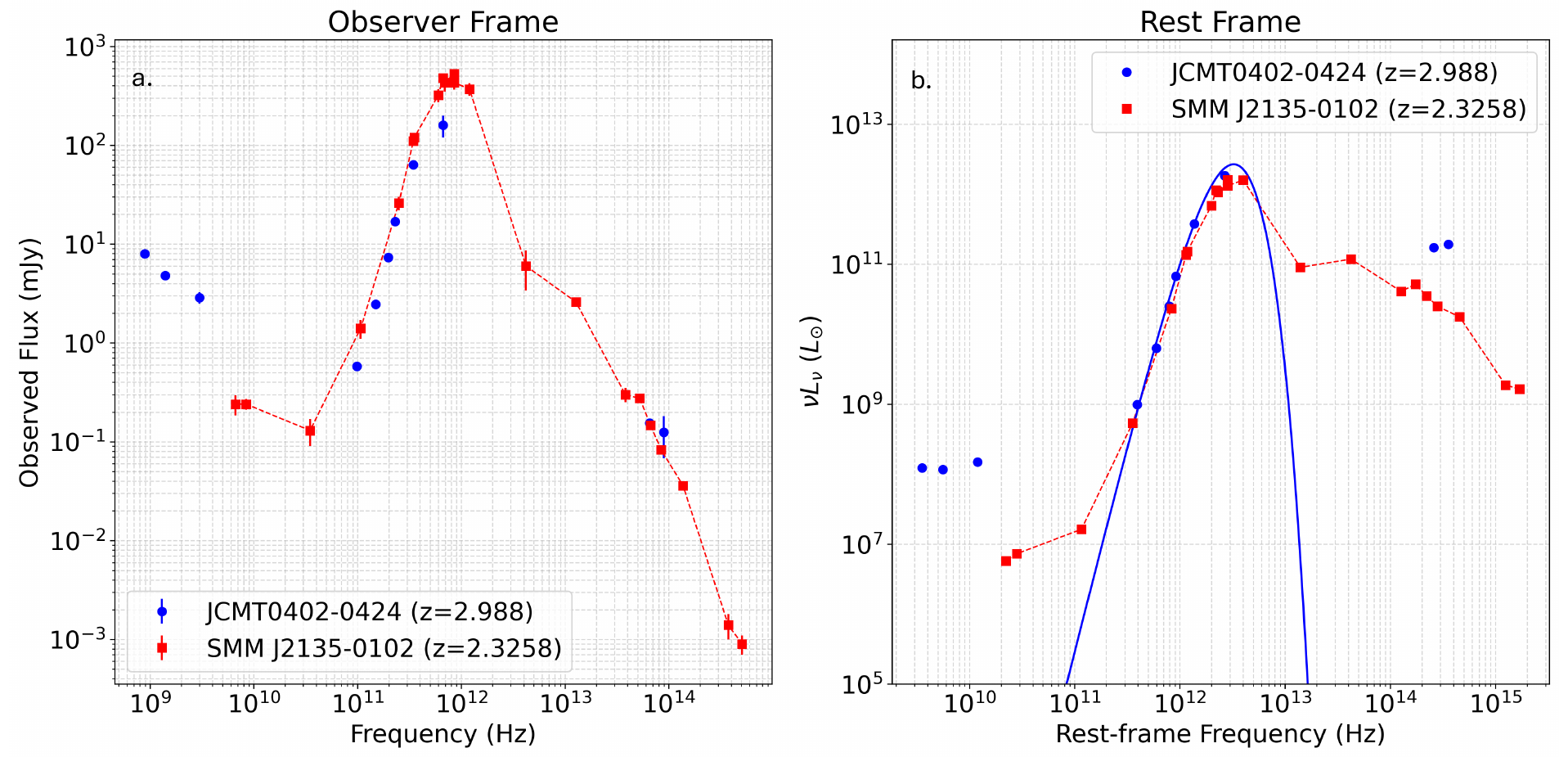}
    \caption{
    \textbf{Broad-band SED of the IC\,210922A counterpart and comparison to a compact-core DSFG.}
\textbf{a,} Observer-frame SED (flux density in mJy) compiled from ALMA, SMA, and JCMT photometry for
JCMT0402$-$0424 at $z=2.988$ (blue points). 
\textbf{b,} Rest-frame SED plotted as $\nu L_\nu$ ($L_\odot$). The blue curve shows the optically-thin
modified-blackbody fit to the far-IR/submm data (WISE excluded), yielding
$T_{\rm dust}=22.66\pm1.70$\,K and $\beta=3.00\pm0.12$. 
For context, red squares (with dashed connector) show the archetypal lensed compact-core DSFG
SMM~J2135$-$0102 at $z=2.3258$. Integrating the best-fit model over 8--1000\,$\mu$m and correcting by the
extended-source magnification $\mu_{\rm ext}=12.20$ gives
$L_{\rm IR}=2.72\times10^{12}\,L_\odot$ ($1.04\times10^{46}$\,erg\,s$^{-1}$), consistent with a ULIRG at $z\approx3$. 
Error bars are $1\sigma$.
}
  \label{fig:fig3}
\end{figure*}
\end{landscape}

\begin{landscape}
\begin{figure}
    \centering
    \includegraphics[width=1.6\textwidth]{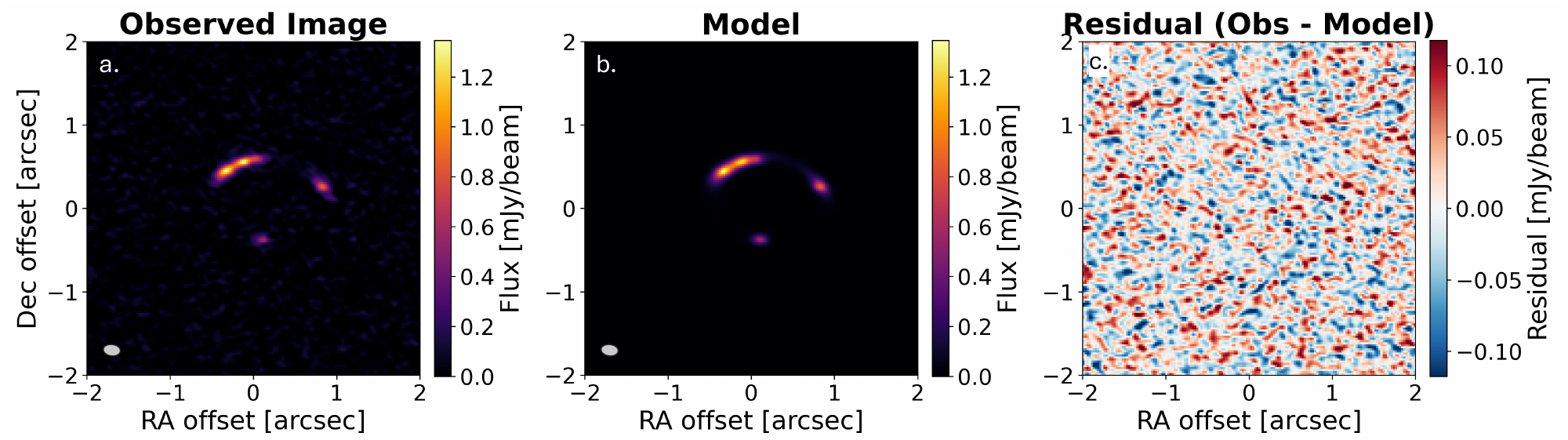}
\caption{
\textbf{ALMA Band~5 (198\,GHz) gravitational lens modeling.}
\textbf{a,} Observed continuum image. 
\textbf{b,} Best-fit model image based on a two-component source model (\textit{Point Source + S\'ersic}).
\textbf{c,} Residual map (observed $-$ model). 
All offsets are relative to the lens center.
The color bars represent the flux density (a, b) and residuals (c), all in mJy\,beam$^{-1}$. 
The synthesized beam is shown as a gray ellipse in the lower-left corner of panels \textbf{a} and \textbf{b}.
The featureless residual map, consistent with the image noise, indicates that the model accurately reproduces the morphology of both the compact and extended components of the lensed source.
}        \label{fig:phase5_paper_figure}
\end{figure}
\end{landscape}

Strong detections of CO(3--2) in Band~3, CO(5--4) in Band~4, and in Band~5 the higher-$J$ CO lines CO(6--5) and CO(7--6) together with [C\,\textsc{i}](2--1) were obtained (Figure \ref{fig:fig2}). 
Each line was detected with high signal-to-noise ratio and individually fit with a Gaussian profile to measure the 
line center (\ref{fig:method_redshift}). 
The observed frequencies are mutually consistent, yielding a robust spectroscopic redshift of $z = 2.9880 \pm 0.0003$, with a statistical precision of $\sigma_z \approx 0.0001$--$0.0005$. 
This provides strong confirmation of the source redshift and demonstrates the utility of simultaneous CO and [C\,\textsc{i}] detections in DSFGs.

Each line profile was further modeled as a combination of Gaussian components 
using non-linear least squares, with the systemic redshift fixed at $z=2.988$ (Figure \ref{fig:fig2}). 
The CO(3--2) line profile is best fit with three components, yielding 
$L'_{\rm CO(3-2)} = 5.26\times10^{10}\ {\rm K\,km\,s^{-1}\,pc^2}$. 
The CO(5--4) line requires a similar three-component model with 
$L'_{\rm CO(5-4)} = 4.61\times10^{10}\ {\rm K\,km\,s^{-1}\,pc^2}$. 
The CO(6--5) line is likewise well described by three components, giving 
$L'_{\rm CO(6-5)} = 1.57\times10^{10}\ {\rm K\,km\,s^{-1}\,pc^2}$. 
The Band~5 CO(7--6)+[C\,\textsc{i}](2--1) spectrum requires three CO(7--6) components and two 
[C\,\textsc{i}](2--1) components, with 
$L'_{\rm CO(7-6)} = 1.13\times10^{10}$ and 
$L'_{\rm [CI](2-1)} = 6.2\times10^{9}\ {\rm K\,km\,s^{-1}\,pc^2}$. 
The narrow components have FWHM $\sim$60--140\,km\,s$^{-1}$, while the broad 
component spans $\sim$390--500\,km\,s$^{-1}$, consistent across bands and 
tracing a multi-component velocity structure.

\subsection*{Optical and Near-Infrared Observations and Analysis}

We obtained archival Subaru/Hyper Suprime-Cam (HSC) imaging data taken in 2016 and 2019 from the SMOKA (Subaru-Mitaka-Okayama-Kiso Archive). The data were processed using the HSC pipeline \citep{Bosch2018}, yielding deep $r$-band imaging that revealed the optical counterpart coincident with the submillimeter source position (Figure~\ref{fig:fig1}c). The galaxy near the geometric center of the lensed configuration coincides with the lens mass centroid, confirming it as the foreground lens.

We also obtained optical imaging with Gemini/GMOS in the $g$, $r$, $i$, and $z$ bands, acquired on 2022 January 8 and archival near-IR imaging from VISTA in the $y$, $J$, and $K_{\rm s}$ bands.
The GMOS data were processed using the DRAGONS pipeline \citep{Labrie2023} and flux-calibrated against the Pan-STARRS1 catalog. Aperture photometry was measured in a $2"$ diameter aperture. The observed magnitudes are:
$g=25.627\pm0.272$, $r=23.632\pm0.039$, $i=22.568\pm0.017$, $z=21.669\pm0.014$, $y=21.421\pm0.404$, $J=19.884\pm0.115$, and $K_{\rm s}=19.504\pm0.179$.

Before SED fitting, we corrected all bands for Galactic extinction following \citep{SchlaflyFinkbeiner2011}; the photometric-redshift analysis used these extinction-corrected fluxes. A WISE counterpart (W1/W2) exists at the same position, but those data were excluded because the WISE PSF cannot cleanly separate the foreground lens from the background DSFG, leading to blended photometry.

We fitted the extinction-corrected SED with \textsc{EAZY}, allowing a non-negative linear combination of templates. The best-fit is obtained with an elliptical-galaxy template at $z_{\rm lens}=1.06\pm0.04$. The redshift posterior $p(z)$ is sharply peaked and essentially unimodal. The SED fit and the corresponding $p(z)$ are shown in \ref{fig:photoz_sed}.

Gemini/GNIRS cross-dispersed near-IR spectroscopy of the lens galaxy was obtained on 2024 February 21 and March 19. The data were reduced using standard IRAF procedures, with 1D spectra extracted from all usable orders. Each order was wavelength-calibrated in \AA, masked in strong telluric regions (1.10--1.17, 1.32--1.50, 1.78--2.02, and $>$2.55\,$\mu$m), and combined after inverse-variance weighting using the variance (VAR) extensions. The result was rebinned to a constant resolving power of $R\simeq1000$ and normalized by the median flux (no absolute flux calibration). \ref{fig:gnirs_spec} shows the final spectrum with a logarithmic flux axis.

Before adopting any redshift prior, we performed a blind, order-by-order search for emission features across the effective GNIRS blue-camera bandpass (0.9--2.5\,$\mu$m). We scanned the spectrum with a sliding Gaussian kernel whose FWHM matched the instrumental resolution (${\rm FWHM}=\lambda/R_{\rm inst}$, with $R_{\rm inst}\simeq1200$), excluded the deepest telluric regions, and required ${\ge}3\sigma$ significance based on the VAR extensions. No convincing line candidates survived this procedure or subsequent visual inspection at any wavelength outside the strong telluric bands, indicating the absence of prominent nebular emission.

These characteristics are consistent with the photometric-redshift solution from the SED fitting, which favors an elliptical template at $z_{\rm phot}=1.06$. Adopting this redshift to quantify the non-detections, we derived $3\sigma$ rest-frame equivalent-width (EW) limits using a Gaussian matched filter and the local $3\sigma$ continuum upper limit. We obtain $EW_{\rm rest}(\mathrm{H}\beta)<2.73$\,\AA\ and $EW_{\rm rest}([\mathrm{O\,III}]\,5007)<1.70$\,\AA. At $z_{\rm phot}=1.06$, [O\,\textsc{ii}]~$\lambda3727$ lies outside the effective blue-camera range (0.768\,$\mu$m) and H$\alpha$ falls within a strong telluric band, so both are unconstrained. These stringent limits---significantly below the typical tens to hundreds of \AA\ seen in $z\sim1$ star-forming galaxies---are fully consistent with a passive (elliptical) lens galaxy at $z\simeq1.06$.

\subsection*{Swift/XRT analysis}

The \textit{Neil Gehrels Swift Observatory} carried out both a six-point tiling of the IceCube-210922A localization \citep{GCN30877} and a pointed observation targeting JCMT0402$-$0424. 
We analysed all \textit{Swift}/XRT photon-counting observations covering JCMT0402$-$0424
(ObsIDs 00014835001, 00014835002, 00014836001, 00014837001 and 00014839001),
obtained between 2021 September 23 and October 2, using the HEASoft package
(version~6.35.2) and the standard \textsc{xrtpipeline} tools.
Events were filtered to the 0.3--10~keV band and grades 0--12. Source counts were extracted from a 20-pixel-radius (47 arcsec) circular region centred on the submillimeter position, with background taken from a nearby source-free region. The combined good exposure time is 4.1~ks and only two counts are recorded in the source region, consistent with the local background. We derive a $3\sigma$ upper limit on the 0.3--10~keV count rate of $2.4\times10^{-3}\,\mathrm{ct\,s^{-1}}$, which corresponds to an absorbed flux limit of $F_{0.3-10\,\mathrm{keV}}\simeq8.5\times10^{-14}\,\mathrm{erg\,cm^{-2}\,s^{-1}}$ (and an unabsorbed flux of $\simeq1.0\times10^{-13}\,\mathrm{erg\,cm^{-2}\,s^{-1}}$) for a power-law spectrum with photon index $\Gamma=2$ and Galactic absorption $N_{\rm H}=6.2\times10^{20}\,\mathrm{cm^{-2}}$.

\subsection*{Gravitational lens modeling}

Gravitational lens modeling was performed using the \textsc{lenstronomy} package \citep{Birrer2018}.

\medskip
\noindent\textbf{Lens identification and mass model.}
The approximate lens center was identified from the geometric distribution of the four lensed images, which form a characteristic Einstein-ring--like configuration. The relative positions and fluxes of the images were measured from the high-resolution ALMA Band~5 continuum data, providing sub-arcsecond astrometric precision. 
The central position coincides with the optical galaxy visible in the Subaru/HSC $r$-band image (Figure~\ref{fig:fig1}c), confirming its identification as the lensing galaxy at $z_{\rm phot}\simeq1.06$. The light profile of the lens was modeled with a S\'ersic profile based on the Subaru/HSC imaging.

The lens mass distribution was described by an elliptical power-law profile with external shear to account for environmental perturbations. Ellipticity and shear components were included to capture the non-axisymmetric contributions to the gravitational potential. The lens equation solver in \textsc{lenstronomy} was used to enforce that all four observed image positions map to a single source-plane position, thereby constraining the lens model parameters directly from the image configuration.

\noindent\textbf{Lens modeling and parameter estimation.}
We tested three classes of source models to reconstruct the lensed background galaxy JCMT0402$-$0424 at $z=2.988$ (\ref{fig:model_comparison}).
A single S\'ersic profile reproduced the global morphology but left significant residuals, while a Chameleon profile yielded modest improvement ($\chi^2/{\rm dof}=1.443$ vs.\ 1.455) but residuals persisted. A double Chameleon model was also attempted but failed to converge due to parameter instability. We therefore adopted a hybrid Point Source + S\'ersic model, which simultaneously accounts for both extended emission and compact features, significantly reducing residuals across all arcs. This configuration provided the best fit to the ALMA imaging, achieving $\chi^2/{\rm dof}=1.397$ and an Akaike Information Criterion (AIC) of 36,195.5, outperforming the S\'ersic (37,708.5) and Chameleon (37,400.4) models. 

Parameter uncertainties were estimated via Markov Chain Monte Carlo (MCMC) sampling in \textsc{lenstronomy}, with 5000 steps and 1000 burn-in iterations after Particle Swarm Optimization (PSO). Convergence was verified using the Gelman--Rubin statistic (\ref{fig:lens_corner}). The best-fit elliptical power-law lens model yields an Einstein radius of $\theta_E = 1.069 \pm 0.006$ arcsec, a mass slope of $\gamma = 1.96^{+0.12}_{-0.10}$, lens ellipticities $e_{1,{\rm lens}}=-0.065\pm0.018$ and $e_{2,{\rm lens}}=-0.071^{+0.007}_{-0.009}$, and external shear components $\gamma_1=0.137^{+0.026}_{-0.022}$ and $\gamma_2=-0.053\pm0.003$. For the background source, the S\'ersic component has an effective radius $R_{\rm eff}=0.064^{+0.007}_{-0.006}$ arcsec ($\sim520$ pc at $z=2.988$), index $n=2.89^{+0.19}_{-0.16}$, ellipticities $e_{1,{\rm src}}=-0.019^{+0.025}_{-0.023}$ and $e_{2,{\rm src}}=-0.163^{+0.021}_{-0.019}$, and magnification $\mu_{\rm ext}=12.2\pm1.5$. The unresolved point source is offset by $\sim50$ mas from the extended centroid with $\mu_{\rm point}=26.9\pm3.5$. The resulting source-plane effective radius of $\sim0.5$ kpc is consistent with compact starburst cores in high-redshift DSFGs, similar to systems such as SDP.81 and SMM\,J2135$-$0102 \citep[e.g.,][]{alma2015,Swinbank2010,Falgarone2017}.

\subsection*{Infrared Luminosity and Star Formation Rate Estimation}

The infrared (IR) luminosity ($L_{\mathrm{IR}}$) and star formation rate (SFR) of the DSFG at $z=2.988$ were derived by fitting its spectral energy distribution (SED). The analysis used millimeter and submillimeter continuum fluxes measured with JCMT/SCUBA-2, SMA, and ALMA, while WISE photometry was excluded. 
The WISE data, at much shorter wavelengths and blended with the foreground lens, 
were excluded because they are far from the far-infrared dust peak and do not provide a reliable AGN diagnostic.
The SED was modeled with a single modified blackbody function,
\[
S_{\nu} \propto \nu^{\beta} B_{\nu}(T),
\]
where $B_{\nu}(T)$ is the Planck function at dust temperature $T$ and $\beta$ the dust emissivity index, under the optically thin approximation. The best-fit parameters are $T = 22.7 \pm 1.7$\,K and $\beta = 3.0 \pm 0.1$, consistent with cold dust in high-redshift star-forming galaxies.

Integrating the model over the rest-frame 8--1000\,$\mu$m range yields an apparent luminosity of $L_{\mathrm{IR,obs}} = (3.3 \pm 0.3)\times10^{13}\,L_{\odot}$. Correcting for the lensing magnification of the extended component ($\mu_{\rm ext}=12.2 \pm 1.5$) gives an intrinsic luminosity of $L_{\mathrm{IR}} = (2.7 \pm 0.3)\times10^{12}\,L_{\odot}$, firmly placing the source in the class of ultra-luminous infrared galaxies (ULIRGs). The intrinsic luminosity corresponds to an SFR of $\sim270$~M$_\odot$\,yr$^{-1}$ for a Chabrier initial mass function (IMF), or $\sim470$~M$_\odot$\,yr$^{-1}$ for a Salpeter IMF, following the \citep{Kennicutt1998} calibration. These values are typical of compact DSFGs at $z\sim3$ \citep[e.g.,][]{Simpson2015,Hodge2016} and consistent with the SED resemblance to the archetypal starburst SMM\,J2135$-$0102 \citep{Swinbank2010}.

\subsection*{Molecular Gas and Dynamical Mass Estimates}

Gaussian decomposition of the observed CO and [C\,I] emission lines in ALMA
Bands~3, 4, and 5 (CO(3--2), CO(5--4), CO(6--5), CO(7--6), and [C\,I](2--1)) was
performed using non-linear least-squares fitting.
Each profile was modeled with multiple Gaussian components, achieving reduced chi-squared values close to unity ($\chi^2_\nu \sim 1$).
The integrated line fluxes
$S\Delta v$ (Jy km s$^{-1}$) were converted into line luminosities following
the standard relation \citep{Solomon1997}:
\begin{equation*}
L' = 3.25 \times 10^7\ 
\frac{S \Delta v\ D_L^2}{(1+z)^3\ \nu_{\rm obs}^2}\ 
\ \ \ [{\rm K\ km\ s^{-1}\ pc^2}],
\end{equation*}
where $D_L$ is the luminosity distance in Mpc, $z$ the redshift, and
$\nu_{\rm obs}$ the observed frequency in GHz. The results are summarized in
Extended Data Table~1.

The molecular gas mass was derived from the CO(3--2) luminosity, converted to
the CO(1--0) equivalent assuming a brightness-temperature ratio of
$r_{31}=L'_{\rm CO(3-2)}/L'_{\rm CO(1-0)}=0.5$
\citep[e.g.,][]{CarilliWalter2013}. The gas mass is then
\begin{equation*}
M_{\rm gas} = \alpha_{\rm CO}\ L'_{\rm CO(1-0)},
\end{equation*}
with a CO-H$_2$ conversion factor $\alpha_{\rm CO}=0.8$
$M_\odot\ ({\rm K\ km\ s^{-1}\ pc^2})^{-1}$ appropriate for compact starburst
galaxies \citep{DownesSolomon1998}. This yields
$M_{\rm gas}\sim 7\times10^9\,M_\odot$ after correcting for lensing.

Independently, the [C\,I](2--1) line was used to estimate the gas mass
\citep{Weiss2003,Papadopoulos2004}. The [C\,I](2--1) luminosity was converted
to [C\,I](1--0) assuming a line ratio $r_{21}=0.6$. The atomic carbon mass is
\begin{equation*}
M_{\rm CI} = 1.902 \times 10^{-4}\ Q(T_{\rm ex})\ e^{23.6/T_{\rm ex}}
\ L'_{\rm CI(1-0)},
\end{equation*}
where $Q(T_{\rm ex})$ is the partition function at excitation temperature
$T_{\rm ex}$. The molecular gas mass then follows from the assumed abundance
$X_{\rm CI}$:
\begin{equation*}
M_{\rm H_2} = \frac{M_{\rm CI}}{6\,X_{\rm CI}}.
\end{equation*}
For $T_{\rm ex}=30$ K and $X_{\rm CI}=3\times10^{-5}$, this yields
$M_{\rm H_2}\sim(4$--$10)\times10^{10}\,M_\odot$.

The dynamical mass was estimated from the observed CO line widths and the
source-plane effective radius inferred from lens modeling
\citep[e.g.,][]{Spilker2015}, using the virial estimator:
\begin{equation*}
M_{\rm dyn} = \frac{5\,R\,\sigma_v^2}{G},
\end{equation*}
where $R$ is the effective radius and $\sigma_v={\rm FWHM}/2.355$ the
velocity dispersion. Adopting the lensing-inferred effective radius of the
dust continuum, $R_{\rm eff}=0.065\pm0.007''$ in the source plane
($0.52\pm0.06$ kpc at $z=2.988$ for a standard cosmology,
$7.9$ kpc arcsec$^{-1}$), the virial estimator yields
$M_{\rm dyn}\sim(1$--$3)\times10^{10}\,M_\odot$ for
FWHM$=400$--$500$ km s$^{-1}$. 
This lensing-inferred radius is consistent with the source model 
and supports the conservative choice of 0.4--0.5 kpc adopted above 
(the CO-emitting region could be modestly more compact than the dust continuum due to
differential magnification and excitation).

The CO-based gas mass and the [C\,I]-based estimate both fall within the
uncertainties of the dynamical mass, indicating that molecular gas
constitutes a substantial fraction of the total mass within the central
$\sim0.5$ kpc.

From the CO-based gas mass ($M_{\rm gas}\!\sim\!7\times10^9\,M_\odot$) 
and the lensing-inferred source-plane radius ($R_{\rm eff}\!\simeq\!0.5$ kpc), 
the mean molecular gas surface density is estimated as 
$\Sigma_{\rm gas}\!\simeq\!M_{\rm gas}/(\pi R_{\rm eff}^2)\!\approx\!9\times10^3\,M_\odot\,{\rm pc^{-2}}$. 
Even for the CO-based estimate, $\Sigma_{\rm gas}$ exceeds $10^3\,M_\odot\,{\rm pc^{-2}}$, 
and increases further if the larger [C\,I]-based mass ($\sim4$--$10\times10^{10}\,M_\odot$) is adopted, consistent with the threshold for the calorimetric condition.

\subsection*{CO Spectral Line Energy Distribution (SLED)}
\label{meth:sled}

\ref{fig:co_sled} compares the CO SLED of JCMT0402$-$0424 with compact starbursts, AGN, and a dust--obscured galaxy (DOG). 
Integrated CO measurements were assembled for SMM~J2135$-$0102 (DSFG) \citep{Danielson2011}, APM~08279+5255 (luminous type-1 quasar; \citep{Greve2009}), H1413+117 (quasar; \citep{Greve2009}), and the compact circumnuclear region of NGC~1068 (nearby Seyfert; \citep{Spinoglio2012}). 
To include a representative DOG with AGN activity, we also use CO observations of W2246$-$0526 at $z\simeq4.6$, for which multi--$J$ CO lines have been measured with ALMA \citep[e.g.][]{Harrington2025}. 
APM~08279+5255, H1413+117, SMM~J2135$-$0102, JCMT0402$-$0424, and W2246$-$0526 all lie at $z\sim2$--5, enabling a direct comparison of CO excitation at broadly similar epochs.
For NGC~1068, we adopt a nuclear template in which hard X-rays from the AGN create an X-ray--dominated region (XDR), typically enhancing high-$J$ CO excitation \citep{Spinoglio2012}. 

By convention, SLEDs are often shown relative to CO(1--0). Because CO(1--0) is not yet available for JCMT0402$-$0424 or W2246$-$0526, two complementary normalizations are presented. 
JCMT0402$-$0424 is represented by ALMA detections of CO(3--2), CO(5--4), CO(6--5), and CO(7--6).
In \ref{fig:co_sled} (left) we show, for each object, the within-object luminosity ratio
$R_J=L'_{\rm CO}(J\!\to\!J\!-\!1)/L'_{\rm CO}(\mathrm{anchor})$, where the anchor transition is chosen as CO(3--2) when available, or otherwise the lowest available transition with $3\!\le\!J_{\rm upper}\!\le\!6$ (the chosen anchor is indicated in the legend; for the DOG W2246$-$0526 this is CO(5--4)). 
When only integrated fluxes $S\Delta v$ are reported, the ratios are converted assuming the standard brightness--temperature scaling, $L' \propto (S\Delta v)/\nu_{\rm obs}^{2}$. 
\ref{fig:co_sled} (right) shows velocity--integrated intensity ratios $S\Delta v(J)/S\Delta v(1\!-\!0)$. 
For JCMT0402$-$0424, CO(1--0) is inferred from CO(3--2) adopting a typical DSFG range 
$r_{31}\equiv L'_{\rm CO(3-2)}/L'_{\rm CO(1-0)}=0.4$--0.7 \citep[e.g.][]{CarilliWalter2013,Bothwell2013}; 
the curve uses the median value ($r_{31}=0.5$) and the vertical error bars reflect this range. 
Since CO(1--0) has not yet been measured for W2246$-$0526, this DOG appears only in the left-hand panel. 
In all cases, error bars in \ref{fig:co_sled} represent published $1\sigma$ measurement uncertainties (or the adopted $r_{31}$ range for JCMT0402$-$0424).

\subsection*{Neutrino luminosity and expected detections from a compact DSFG}

A compact, high-$z$ DSFG associated with IC\,210922A is considered under a calorimetric starburst scenario. In a compact core with very high gas surface density ($\Sigma_{\rm gas}\!\gtrsim\!10^3~M_\odot\,{\rm pc^{-2}}$), cosmic-ray protons accelerated in the compact starburst environment are efficiently confined and lose a large fraction of their energy via inelastic $pp$ collisions, producing pions that decay into neutrinos and $\gamma$-rays.

The neutrino spectral normalization is tied to the \emph{bolometric} neutrino energy flux at Earth, rather than to a detector-dependent integration window. From the infrared luminosity of the DSFG ($L_{\rm IR}=1.0\times10^{46}\,{\rm erg\,s^{-1}}$), adopting a fiducial effective conversion factor $\eta_{\rm CR}\!\approx\!0.1$ for the compact starburst core and a neutrino branching factor of $\simeq 1/3$, the bolometric neutrino luminosity is $L_\nu \simeq \tfrac{1}{3}\eta_{\rm CR}L_{\rm IR} \approx 3.3\times10^{44}\,{\rm erg\,s^{-1}}$.
%
Here, $\eta_{\rm CR}$ is treated as an effective phenomenological conversion factor for the compact starburst core, rather than as a quantity attributable exclusively to isolated supernova remnants. In such dense environments, additional or collective acceleration channels---including stellar-wind shocks, superbubbles, starburst-driven outflows, and turbulent or multiple-shock reacceleration---may also contribute to the non-thermal particle budget, so the adopted normalization should be regarded as physically motivated but not uniquely tied to a supernova-only estimate  \citep{RuszkowskiPfrommer2023}. 
The observed cm-wave radio emission shown in Figure~\ref{fig:fig3} is qualitatively consistent with the presence of enhanced non-thermal activity in such an environment, although quantifying its contribution to the proton power would require dedicated radio-SED and particle-transport modeling and is beyond the scope of the present work \citep{Tabatabaei2025}.

For comparison, a conservative estimate based on the supernova power budget gives a lower neutrino luminosity than our fiducial compact-core normalization. Adopting ${\rm SFR}\sim 400~M_\odot\,{\rm yr^{-1}}$ and one core-collapse supernova per $\sim 50~M_\odot$ of newly formed stars implies a supernova rate of $\sim 8~{\rm yr^{-1}}$, corresponding to a total supernova mechanical power of $\sim 2.5\times10^{44}~{\rm erg\,s^{-1}}$. If a canonical $\sim 10\%$ fraction of this mechanical power is transferred to cosmic-ray protons, the resulting cosmic-ray power is $\sim 2.5\times10^{43}~{\rm erg\,s^{-1}}$, implying a characteristic neutrino luminosity of order $\sim 10^{43}~{\rm erg\,s^{-1}}$ for a hadronic branching factor of $\sim 1/3$ \citep{Caprioli2012}. This is approximately one order of magnitude below our fiducial phenomenological estimate above. We therefore interpret the adopted $\eta_{\rm CR}\approx0.1$ as a compact-core effective normalization rather than a unique supernova-only estimate, and regard the supernova-budget estimate as a conservative lower-end benchmark. This interpretation is also broadly consistent with empirical FIR--$\gamma$-ray correlations and related modeling of star-forming galaxies, which indicate that highly star-forming systems can approach the calorimetric regime while still exhibiting substantial source-to-source and transport-dependent scatter \citep{Ajello2020,Kornecki2020,Werhahn2021}.

Using the luminosity distance at $z\simeq3$ ($D_L\simeq8.0\times10^{28}$ cm), the unlensed bolometric neutrino flux is $F_{\nu,{\rm bol}}^{(0)}=L_\nu/(4\pi D_L^2)$. High-energy neutrinos follow null geodesics analogously to photons, so gravitational lensing magnifies the observed flux by a factor $\mu_\nu$, giving $F_{\nu,{\rm bol}}^{\rm obs}=\mu_\nu\,F_{\nu,{\rm bol}}^{(0)}$.

A single power-law differential spectrum at Earth is adopted, $\phi_\nu(E)=K_{\rm obs}\,E^{-\gamma}$ (per flavor unless noted). The normalization $K_{\rm obs}$ is determined once from $F_{\nu,{\rm bol}}^{\rm obs}$ over a fixed bolometric band representative of $pp$-produced neutrinos in the source frame ($E_{\rm bol,min}^{\rm src}=1~{\rm GeV}$ to $E_{\rm bol,max}^{\rm src}=1~{\rm EeV}$), mapped to the observer frame by $(1+z)$. The expected number of detected events over an exposure $T$ is then
\[
N_\nu \;=\; T \int \phi_\nu(E)\,A_{\rm eff}(E)\,dE,
\]
with $A_{\rm eff}(E)$ taken from the public 7-year IceCube point-source effective areas \citep{Aartsen2017}. Because $K_{\rm obs}\propto F_{\nu,{\rm bol}}^{\rm obs}\propto \mu_\nu$, the event expectation scales linearly with magnification ($N_\nu\propto\mu_\nu$). The fraction of bolometric \emph{energy} in an analysis band $[E_1,E_2]$ is
\[
f_{\rm bol}(E_1,E_2;\gamma)=\frac{\int_{E_1}^{E_2}E^{1-\gamma}dE}{\int_{E_{\rm bol,min}^{\rm obs}}^{E_{\rm bol,max}^{\rm obs}}E^{1-\gamma}dE},
\]
ensuring that $N_\nu$ is non-decreasing as the integration band widens or as $A_{\rm eff}$ increases.

For $z=3.0$ and $\mu_\nu=1$ (unlensed), the 10-year expectations (per flavor) are:
$\gamma=2.0$: $N_\nu(1$--$10^3~{\rm TeV})=1.9\times10^{-3}$ and $N_\nu(0.1$--$10^6~{\rm TeV})=2.2\times10^{-3}$;
$\gamma=2.8$: $N_\nu(1$--$10^3~{\rm TeV})=5.3\times10^{-6}$ and $N_\nu(0.1$--$10^6~{\rm TeV})=7.4\times10^{-6}$.
Three lensing scenarios are relevant for JCMT0402$-$0424, reflecting the emission geometry relative to the ALMA morphology. In an unlensed/intrinsic scenario ($\mu_\nu=1$), the values above apply directly. If neutrino emission is co-spatial with the kpc-scale S\'ersic component (“extended” case; $\mu_\nu=12.2$), the full-range (0.1--$10^6$~TeV) expectations scale to $N_\nu\simeq2.7\times10^{-2}$ for $\gamma=2.0$ and $N_\nu\simeq9.1\times10^{-5}$ for $\gamma=2.8$. If emission is dominated by a compact core co-spatial with the unresolved ALMA component (“point-like” case; $\mu_\nu=26.9$), the same band gives $N_\nu\simeq5.9\times10^{-2}$ ($\gamma=2.0$) and $N_\nu\simeq2.00\times10^{-4}$ ($\gamma=2.8$). Characteristic intervals scale inversely with $\mu_\nu$; thus, the unlensed $\sim4.6\times10^3$ yr per event for $\gamma=2.0$ reduces to $\sim3.7\times10^2$ yr (extended) and $\sim1.7\times10^2$ yr (point-like). 
Even in the magnified compact-core case, the 10-year expectation per flavor remains $N_\nu \ll 1$, so repeated detections from any individual compact DSFG are not anticipated within the current IceCube observation, consistent with non-detections in follow-up searches \citep{GCN30872}.



\subsection*{Diffuse neutrino background modeling}

The diffuse, all-sky \emph{per-flavor} neutrino intensity $E^2\Phi(E)$ is obtained by convolving infrared (IR) luminosity functions (LFs) with a source-frame neutrino spectrum and the cosmological volume element. Two emitter populations are included: (i) dusty star-forming galaxies (DSFGs) with a two-component morphology (compact core and extended\,+\,LIRG), and (ii) obscured AGN (DOGs). For comparison with IceCube, all curves are shown per flavor ($\nu_e\!:\!\nu_\mu\!:\!\nu_\tau=1\!:\!1\!:\!1$ at Earth); the all-flavor prediction is larger by a factor of 3.

\medskip
\noindent\textbf{Luminosity functions and redshift evolution.}
High-resolution ALMA surveys indicate that the DSFG population contains both compact-core systems with effective radii of $\lesssim$kpc and more extended galaxies \citep[e.g.,][]{Simpson2015,Hodge2016,Fujimoto2017}. Compact-core DSFGs are predominantly starburst-driven (cool-dust selection), whereas mid-IR selected dusty galaxies (DOGs) show a higher AGN fraction \citep[e.g.,][]{Toba2015,Toba2017}. Because the particle-acceleration environments differ between starburst-dominated DSFGs and AGN-dominated DOGs, these populations are treated separately.

For DSFGs, a Casey/Zavala-type LF with an evolving knee $L_\star(z)$ and normalization $\phi_\star(z)$ is adopted \citep{Casey2018,Zavala2021}, Schechter-like at the bright end \citep{Schechter1976}. Below/above $L_\star$ the adopted slopes are $\alpha_{\rm LF}=-0.42$ and $\beta_{\rm LF}=-3.0$, with $\log_{10}L_{\star,0}=11.1$, $\log_{10}\phi_{\star,0}=-3.5$, and $z_{\rm turn}=2.0$. 
The high-$z$ evolution follows $\phi_\star\propto(1+z)^{\psi_2}$. A global LF scale factor, $\mathrm{flux\mbox{-}scale}$, is introduced as an effective occupancy factor for the neutrino-active DSFG subset, i.e. a population-level rescaling rather than a change in the spectral shape.
The DSFG LF is divided into compact and extended components using a redshift-dependent compact fraction $f_c(z)$ guided by ALMA results. Two functional forms are considered and give consistent totals within uncertainties:
\[
f_c^{\rm Gauss}(z)=\max\!\left[\mathrm{gfloor},\,\mathrm{gpeak}\,\exp\!\left(-\frac{(z-\mathrm{gz0})^2}{2\,\mathrm{gsig}^2}\right)\right],\quad
f_c^{\rm logi}(z)=f_{\min}+\frac{f_{\max}-f_{\min}}{1+\exp\!\left(-\frac{z-\mathrm{z_t}}{\Delta z}\right)}.
\]
We do not assume that all DSFGs are mergers or compact neutrino-efficient systems at all times. Rather, the compact subset is interpreted as an effective neutrino-bright phase within the broader DSFG population, plausibly associated with merger-driven compaction or similarly extreme, short-lived conditions near coalescence or immediate post-coalescence, while allowing for substantial morphological diversity in the parent DSFG population \citep{Gillman2024,Ellison2025}.
DOGs are described with a double--power-law LF appropriate for an AGN-dominated bright end \citep{Toba2015,Toba2017},
\[
\phi(L)=\frac{\phi_\star}{(L/L_\star)^{\alpha}+(L/L_\star)^{\beta}},
\]
modulated by a Gaussian redshift envelope $g(z)=\exp[-(z-z_0)^2/(2\sigma_z^2)]$ and an amplitude factor $\mathrm{dog\mbox{-}scale}$ that reflects the duty cycle and stringent mid-IR color cuts.

\medskip
\noindent\textbf{Source-frame spectrum and the choice of SBPL.}
Each source spectrum is represented by a smoothly broken power law (SBPL; \citep{Massaro2004,Abdo2009,IceCube2015}) with smoothness $s=2$, optionally with an exponential cut-off \citep{Kelner2006}. For the \emph{compact DSFG core}, an SBPL is favored over a single cut-off power law because: (i) the lens reconstruction implies at least two compact subcomponents (an unresolved hotspot plus a compact S\'ersic core), whose superposition with slightly different maximum energies/grammage produces a broad, gradual steepening; (ii) dense starburst cores are expected to transition from advection-dominated calorimetry to diffusion-dominated escape with increasing rigidity, yielding a smooth slope change $\Delta\gamma\!\sim\!\delta$; and (iii) the per-flavor IceCube spectrum shows a gentle shoulder across $E_\nu\!\sim\!30$--300\,TeV that is naturally reproduced by an SBPL with minimal tuning.  
For the \emph{extended\,+\,LIRG} component, where escape dominates in more diffuse gas, a single unbroken power law suffices within current uncertainties. \emph{No low-energy calorimetric suppression factor $f_{\rm cal}$ is applied} so as not to duplicate transport-induced softening already encoded by the SBPL. The per-flavor neutrino luminosity is linked to the CR injection power via an efficiency $\eta$ (all-flavor $L_{\nu,{\rm all}}\simeq0.5\,L_{\rm CR}$, per-flavor $L_{\nu,1{\rm fl}}\simeq0.17\,L_{\rm CR}$ in the calorimetric limit), with distinct $\eta$ values for ULIRGs and LIRGs.

\medskip
\noindent\textbf{Cosmological forward model and parameter choices.}
For observed energy $E$, a source at redshift $z$ is sampled at $E'=(1+z)E$. The per-flavor differential intensity is
\[
\Phi(E)=\frac{1}{4\pi}\int_{0}^{z_{\max}}\!\!dz\,\frac{dV_c}{dz\,d\Omega}\!
\int_{\log_{10}L_{\min}}^{\log_{10}L_{\max}}\!\!d\log_{10}L\;
\phi(L)\,g(z)\,\frac{1}{(1+z)}\,\frac{1}{4\pi D_L^2(z)}\,
\left.\frac{dL_{\nu,{\rm 1fl}}}{dE'}\right|_{E'=(1+z)E}\!,
\]
and $E^2\Phi(E)$ is plotted. The integration adopts $E\in[0.1,10^{6}]$\,TeV (observed) and $z\in[0,7]$; luminosity limits are $L\in[10^{11},3\times10^{13}]\,L_\odot$ for DSFGs (LIRG\,+\,ULIRG) and $L\in[10^{12},10^{14}]\,L_\odot$ for DOGs. SBPL normalization uses logarithmic-energy weighting to conserve $L_{\nu,{\rm 1fl}}$ over $[\ln E'_{\min},\ln E'_{\max}]$. 

\begin{figure}
  \centering
  \includegraphics[width=\textwidth]{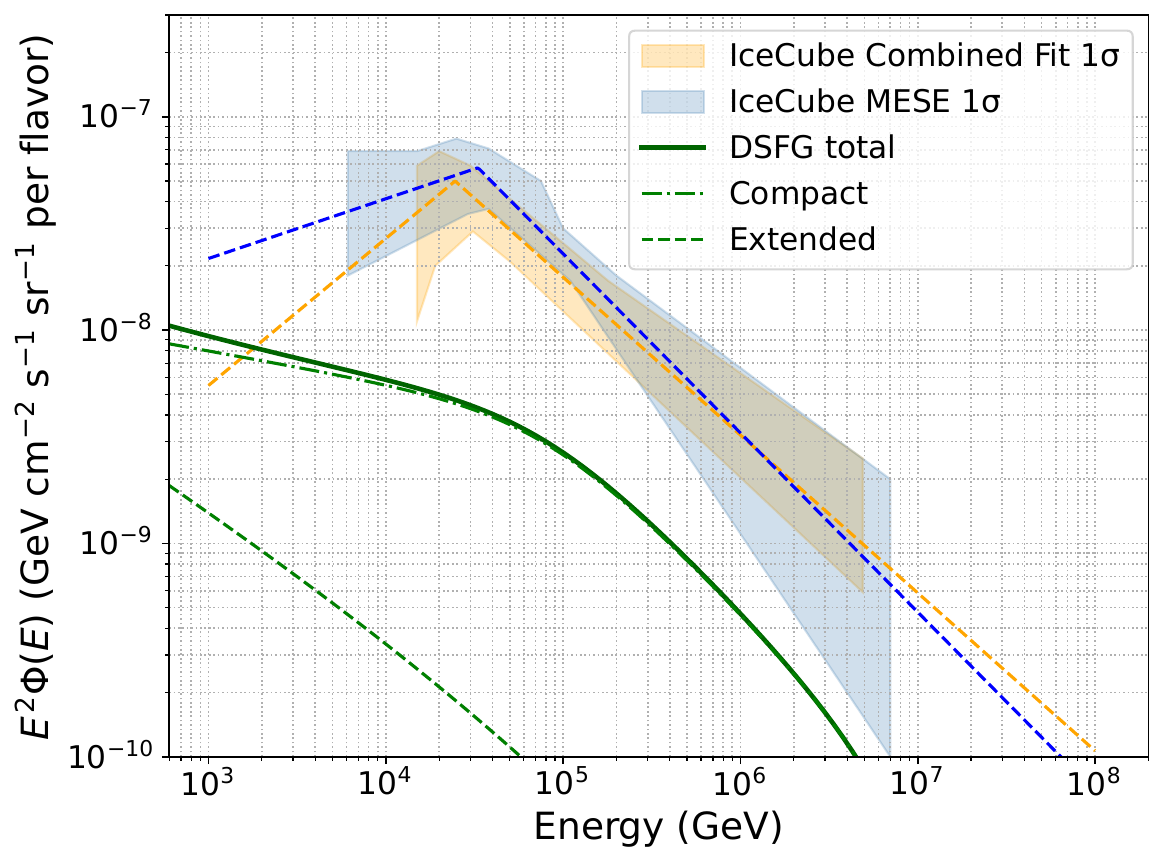}
\vspace{1ex}
\caption{
\textbf{Diffuse neutrino flux from DSFGs compared to IceCube spectra.}
Compact-core DSFGs (dash-dotted green) and extended DSFGs (dashed green) are shown separately, with their sum in dark green. 
For comparison, the IceCube best-fit broken power-law spectra are overplotted (orange dashed: Combined Fit, blue dashed: MESE), with shaded bands denoting the 1$\sigma$ uncertainties\citep{Abbasi2025}. 
The vertical axis shows the per-flavor flux $E^{2}\Phi(E)$. 
\textbf{The compact-core component provides the dominant share of the DSFG contribution around 10--100\,TeV, while the extended component contributes mainly at lower energies.}
}

  \label{fig:dsfg_only}
\end{figure}

Unless stated otherwise, the SBPL smoothness is $s=2$ and no low-energy calorimetric suppression ($f_{\rm cal}$) is applied. 
The compact ULIRG core is assigned a separate per-source normalization factor, $\mathrm{compact\mbox{-}scale}$, which rescales the fiducial calorimetric compact-core neutrino luminosity without altering the source spectral form; by contrast, $\mathrm{flux\mbox{-}scale}$ acts at the population level through the LF normalization. 
In the reference SBPL model used for Figure~\ref{fig:dsfg_only}, we adopt $\mathrm{flux\mbox{-}scale}=0.10$, $\mathrm{compact\mbox{-}scale}=0.10$, $L_{\min}=10^{11}L_\odot$, $\eta_{\rm LIRG}=5\times10^{-4}$, $\eta_{\rm ULIRG}=10^{-3}$, and $\psi_2=-3.5$. The compact DSFG core uses $\gamma_1=2.15$, $\gamma_2=2.85$, a break at $E_b=8.0\times10^{4}$\,GeV (observed frame), and an exponential cut-off with $E_{\rm cut}=1.0\times10^{7}$\,GeV and $k=1.5$; the extended\,+\,LIRG component uses $\gamma_1=2.55$, $\gamma_2=2.85$, $E_b=3.0\times10^{4}$\,GeV, and $E_{\rm cut}=1.5\times10^{6}$\,GeV with $k=1.5$. The compact fraction $f_c(z)$ is evaluated in both Gaussian and logistic forms as described above.

\ref{fig:dsfg_plus_agn} additionally shows the DSFG\,+\,DOG comparison and a soft cut-off variant of the DSFG spectrum. In that variant, the population-level and LF parameters are kept unchanged, while the compact component adopts a softer high-energy attenuation with $E_{\rm cut}=10^{6}$\,GeV and $k=0.7$; the extended component is unchanged. This choice preserves the overall cosmological framework while illustrating the sensitivity of the highest-energy behavior to the compact-core spectral cut-off.


\medskip
\noindent\textbf{Gamma-ray background check.}
As a consistency check, we convert the modeled DSFG neutrino output into an effective escaping $\gamma$-ray contribution and compare it with the non-blazar residual of the extragalactic $\gamma$-ray background (EGB) above 50\,GeV \citep{Ackermann2016EGB50,Ackermann2015IGRB}. For each source, the modeled per-flavor neutrino spectrum is integrated over $E_\nu=10^{5}$--$10^{8}$\,GeV (observed frame) to obtain the neutrino energy flux, converted to all flavor by a factor of three, and then mapped to an electromagnetic energy budget assuming $k_{\rm em/\nu}=1$. The escaping $\gamma$-ray component is parameterized by a band-averaged escape fraction $\mathrm{egb\mbox{-}fesc}$ and distributed as a flat $E^2\Phi_\gamma$ cascade over 50--820\,GeV. This check is applied only to the $\gamma$-ray consistency estimate and does not modify the neutrino spectra shown in Figures~\ref{fig:dsfg_only} and \ref{fig:dsfg_plus_agn}. This observed-frame 50--820\,GeV comparison is intended as a band-averaged consistency check for the IceCube-relevant neutrino component, rather than as a full source-frame prediction of the low-energy $\gamma$-ray spectrum. A more complete treatment would require explicit modeling of the source-frame electromagnetic cascade, cosmological redshifting, and EBL propagation over the full DSFG/DOG population. For simplicity and direct comparison with the Fermi-LAT residual above 50\,GeV, we perform this consistency check in terms of the integrated photon flux rather than the band-integrated energy flux.

For the reference SBPL configuration we adopt $\mathrm{egb\mbox{-}fesc}=0.8$, corresponding to a mildly suppressed escaping 50--820\,GeV component from compact, dust-obscured starburst cores, while avoiding the extreme assumption of a fully $\gamma$-dark source. The resulting integrated DSFG $\gamma$-ray intensity is compared with the non-blazar residual above 50\,GeV, estimated as
\[
I_{\gamma,\mathrm{allow}}=(1-q_{\rm blazar})\,I_{\rm EGB}(>50\,{\rm GeV}),
\]
with $I_{\rm EGB}(>50\,{\rm GeV})=2.4\times10^{-9}\ {\rm cm^{-2}\ s^{-1}\ sr^{-1}}$ and $q_{\rm blazar}=0.86$ \citep{Ackermann2016EGB50}. Under this prescription, the reference DSFG model remains consistent with the residual EGB budget while still providing a subdominant contribution to the diffuse neutrino background. A more conservative upper-limit test with $\mathrm{egb\mbox{-}fesc}=1$ yields a correspondingly tighter but still near-boundary constraint.

We note that the DOG contribution shown in \ref{fig:dsfg_plus_agn} is included only for the neutrino spectral comparison and is not part of the DSFG EGB consistency check. If the neutrino production in DOGs is powered by deeply obscured AGN activity, the accompanying gamma rays may be strongly attenuated and reprocessed to lower energies, potentially into the MeV band, before escape \citep{Murase2020b}.

\medskip
\noindent\textbf{Reference IceCube spectra.}

The per-flavor broken-power-law curves overplotted in Figures~\ref{fig:dsfg_only} and \ref{fig:dsfg_plus_agn} use the Combined-Fit and MESE parameters from \cite{Abbasi2025}, with $(\phi_0,\gamma_1,\gamma_2,\log_{10}E_b/{\rm GeV})=(1.77\times10^{-18},1.31,2.74,4.39)$ and $(2.28\times10^{-18},1.72,2.84,4.52)$, respectively, pivoted at $E_0=10^5$\,GeV.

\section*{Data Availability}

The observational data supporting this study are publicly available from the respective observatory archives.
JCMT/SCUBA-2 data are available from the Canadian Astronomy Data Centre (CADC) under program ID M21BP064.
SMA data are available from the SMA data archive under project ID 2021A-A001.
ALMA data are available from the ALMA Science Archive under project code 2023.1.01229.S.
Subaru/HSC data are available from the SMOKA archive under observation IDs o16319 and o18227.
Gemini data are available from the Gemini Observatory Archive under program IDs GN-2024A-FT-202 and GN-2021B-Q-117.
VISTA data are available from the ESO Science Archive under program ID 179.A-2010(N).
Swift/XRT data are available from HEASARC under ObsIDs 00014835001, 00014835002, 00014836001, 00014837001 and 00014839001.

\section*{Code Availability}

JCMT/SCUBA-2 data were reduced with the \textsc{Starlink/SMURF} package \citep{Chapin2013}, 
SMA data with the \textsc{MIR} pipeline followed by imaging in \textsc{MIRIAD} \citep{MIRIAD}, 
and ALMA data with the CASA pipeline \citep{casa1,casa2}.
Subaru/HSC data were processed with the HSC pipeline \citep{Bosch2018}, 
Gemini/GMOS imaging with \textsc{DRAGONS} \citep{Labrie2023}, 
and GNIRS spectroscopy with standard \textsc{IRAF} procedures. 
HEASoft is available from the HEASARC website
(\url{https://heasarc.gsfc.nasa.gov/lheasoft/}).
%
Photometric-redshift fitting was performed using \textsc{EAZY} \citep{Brammer2008}, 
and gravitational lens modeling with the publicly available \textsc{lenstronomy} package \citep{Birrer2018}. 
Custom Python scripts used for additional analysis, modelling and figure generation are available from the corresponding author upon reasonable request.

\section*{Acknowledgements}

We thank the staff of the James Clerk Maxwell Telescope (JCMT), the Submillimeter Array (SMA), the Atacama Large Millimeter/submillimeter Array (ALMA), and the Gemini Observatory for their support during the observations. This research has made use of data from the Canadian Astronomy Data Centre (CADC), the Subaru--Mitaka--Okayama--Kiso Archive (SMOKA), the Gemini Observatory Archive, and the ESO Science Archive Facility. 

The James Clerk Maxwell Telescope is operated by the East Asian Observatory on behalf of the National Astronomical Observatory of Japan, Academia Sinica Institute of Astronomy and Astrophysics, the Korea Astronomy and Space Science Institute, the National Astronomical Research Institute of Thailand, and the Center for Astronomical Mega-Science, with additional funding from the Science and Technology Facilities Council (UK) and participating universities in the United Kingdom and Canada. 

 The Submillimeter Array is a joint project between the Smithsonian Astrophysical Observatory and the Academia Sinica Institute of Astronomy and Astrophysics, and is funded by the Smithsonian Institution and Academia Sinica. This paper makes use of the following ALMA data: ADS/JAO.ALMA\#2023.1.01229.S. ALMA is a partnership of ESO (representing its member states), NSF (USA), and NINS (Japan), together with NRC (Canada), NSTC and ASIAA (Taiwan), and KASI (Republic of Korea), in cooperation with the Republic of Chile. The Joint ALMA Observatory is operated by ESO, AUI/NRAO, and NAOJ. 

 Based in part on observations obtained at the international Gemini Observatory, a program of NSF NOIRLab, which is managed by the Association of Universities for Research in Astronomy (AURA) under a cooperative agreement with the U.S. National Science Foundation on behalf of the Gemini Observatory partnership: the U.S. National Science Foundation (United States), National Research Council (Canada), Agencia Nacional de Investigaci\'{o}n y Desarrollo (Chile), Ministerio de Ciencia, Tecnolog\'{i}a e Innovaci\'{o}n (Argentina), Minist\'{e}rio da Ci\^{e}ncia, Tecnologia, Inova\c{c}\~{o}es e Comunica\c{c}\~{o}es (Brazil), and Korea Astronomy and Space Science Institute (Republic of Korea). Part of the data (GNIRS observations) were obtained via the Subaru--Gemini Time Exchange Program. The Subaru Telescope is operated by the National Astronomical Observatory of Japan, and data were also obtained from SMOKA, operated by the Astronomy Data Center, NAOJ.

This research has made use of data and/or software provided by the High Energy Astrophysics Science Archive Research Center (HEASARC), which is a service of the Astrophysics Science Division at NASA/GSFC.
We thank Dr. Amy Lien for providing the Swift/BAT data and for valuable assistance in their interpretation.
B.H., S.S.K. and Y.Ma. acknowledge support from JSPS KAKENHI Grant Numbers JP23K03449, JP23H04899, JP26K00733 and JP26K00750. S.S.K. also acknowledges support from the Tohoku Initiative for Fostering Global Researchers for Interdisciplinary Sciences (TI-FRIS).

\section*{Author Contributions}
Y.U. and K.H. initiated the project and coordinated the overall study. K.H. led the ALMA observations as PI, and Y.U. led the JCMT, SMA and Gemini observing programs as PI. Y.U. led the data analysis, interpretation and manuscript drafting, with substantial contributions from K.H. Contributions to discussions of the ALMA follow-up strategy, observational methodology and interpretation of the results were provided by B.H., S.S.K., Y.Ma., Y.Mi., H.N. and K.N. M.K. and R.S. provided the Gemini data and contributed to the multi-wavelength context. All authors reviewed the manuscript.

\section*{Competing Interests Statement}

The authors declare no competing interests.

\clearpage


\clearpage

\clearpage

\clearpage

\clearpage

\clearpage

\section*{Extended Data}

\begin{extfigure}[!bt]
  \centering
  \includegraphics[width=0.92\linewidth]{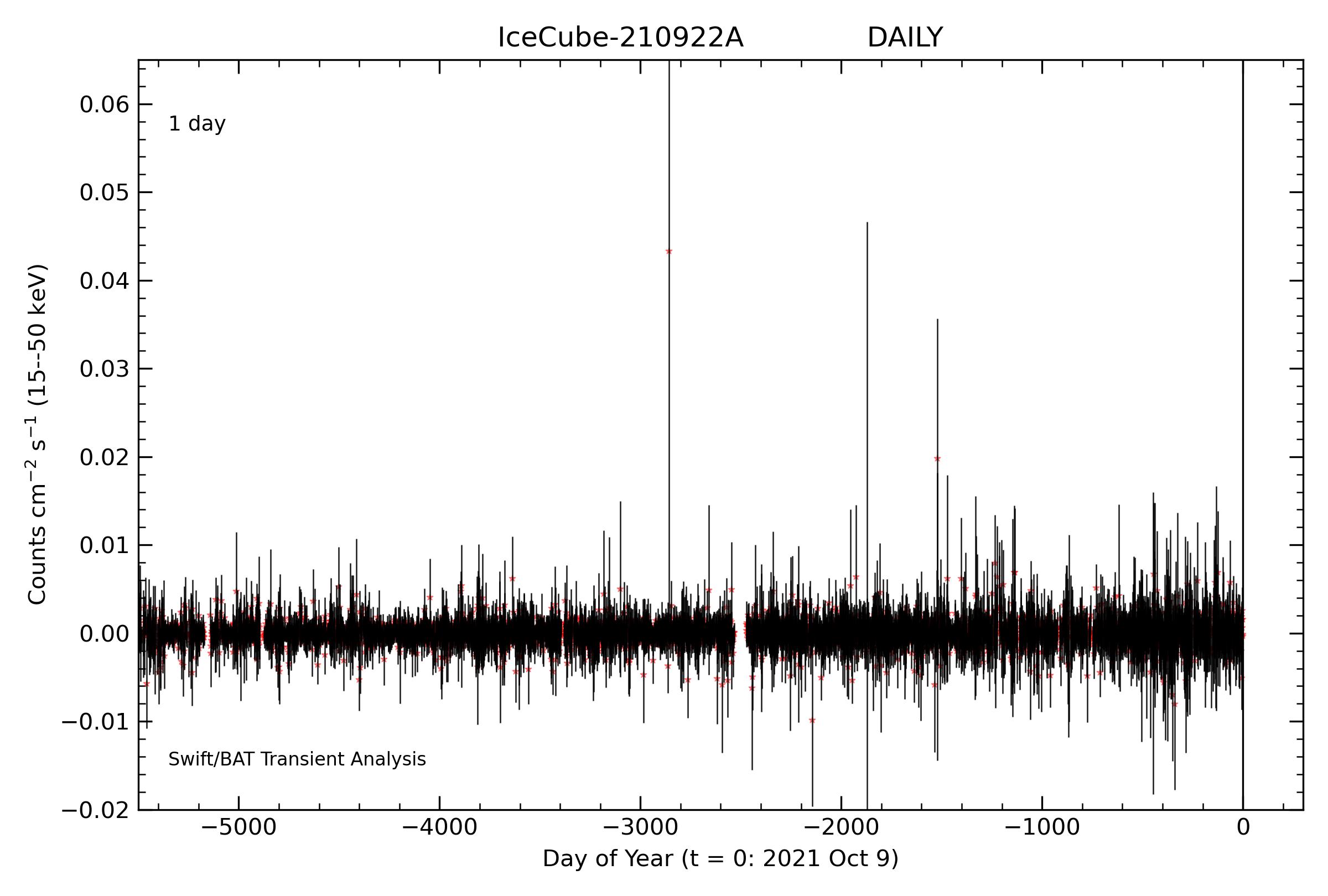}
  \vspace{1em}
  \parbox{0.92\linewidth}{\small
    \vspace{1em}
\textbf{Extended Data Figure \arabic{extfig} | Swift/BAT long-term hard X-ray monitoring at the IC\,210922A position.}
Daily 15--50\,keV count-rate light curve from the BAT Transient Monitor\cite{Krimm}, covering $\gtrsim 5000$ days prior to the IceCube alert. 
Red points show the daily measurements with black error bars indicating $1\sigma$ uncertainties (units: counts\,cm$^{-2}$\,s$^{-1}$). 
The count rates scatter around zero with no evidence for persistent excess emission or significant flares, excluding the presence of a bright variable hard X-ray source such as an AGN within the IceCube localization. }
  \label{fig:swift_bat_monitor}
\end{extfigure}

\clearpage
\begin{landscape}
\begin{extfigure}[t]
  \centering
  \includegraphics[width=0.95\linewidth]{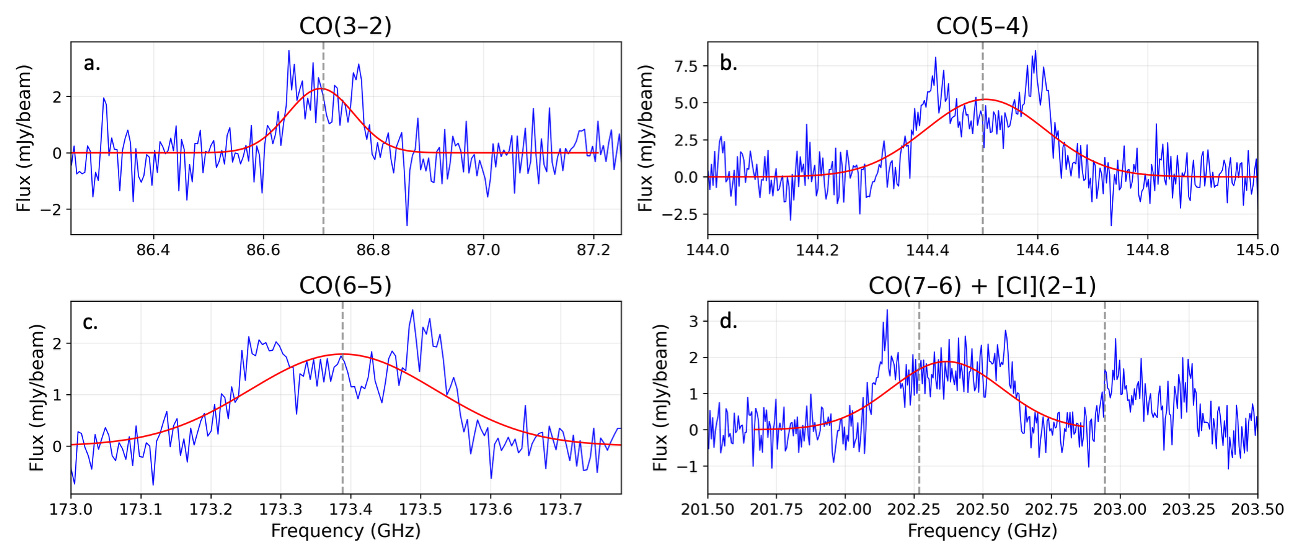}
  \vspace{1em}
  \parbox{0.95\linewidth}{\small
    \vspace{1em}
\textbf{Extended Data Figure \arabic{extfig} | Redshift determination from single-Gaussian line fits.} 
\textbf{a,} CO(3--2) spectrum observed in ALMA Band~3. 
\textbf{b,} CO(5--4) spectrum in Band~4. 
\textbf{c,} CO(6--5) spectrum in Band~5. 
\textbf{d,} CO(7--6) together with [C\,\textsc{i}](2--1) in Band~5. 
In each panel, the observed spectrum is shown in blue and the single-Gaussian fit in red. 
Comparing the fitted observed frequencies with the known rest frequencies yields consistent redshifts from all lines, which provides a robust estimate of $z = 2.9880 \pm 0.0003$; the line-center errors correspond to a statistical precision of $\sigma_z \simeq (1$--$5)\times10^{-4}$ depending on the transition. 
This multi-line identification securely establishes the spectroscopic redshift of the DSFG.
}
  \label{fig:method_redshift}
\end{extfigure}
\end{landscape}

\begin{extfigure}
  \centering
  \includegraphics[width=\linewidth]{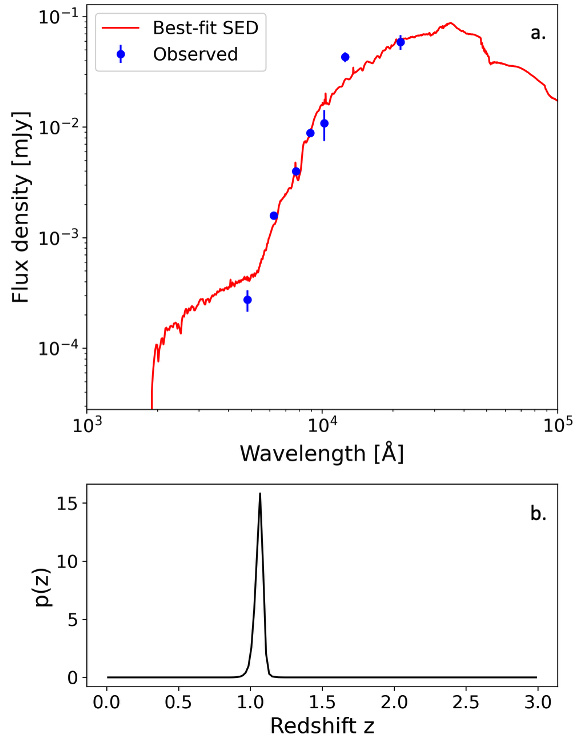}
  \vspace{1em}
  \parbox{\linewidth}{\small
    \vspace{1em}
\textbf{Extended Data Figure \arabic{extfig} | Photometric redshift fit for the lens galaxy.}
\textbf{a,} Broadband SED built from Gemini/GMOS ($gri z$) and VISTA ($yJK_{\rm s}$) photometry measured in a $2\arcsec$ aperture. 
The SED shown and used in the fit has been corrected for Galactic extinction following \citep{SchlaflyFinkbeiner2011}. 
Blue points show the observed fluxes with $1\sigma$ errors, and the red curve is the \textsc{EAZY} best-fit template (elliptical galaxy).  
\textbf{b,} Photometric-redshift probability distribution, yielding $z_{\rm lens}=1.06\pm0.04$. 
WISE W1/W2 were excluded due to blending with the background DSFG.}
  \label{fig:photoz_sed}
\end{extfigure}

\begin{extfigure}
  \centering
  \includegraphics[width=\linewidth]{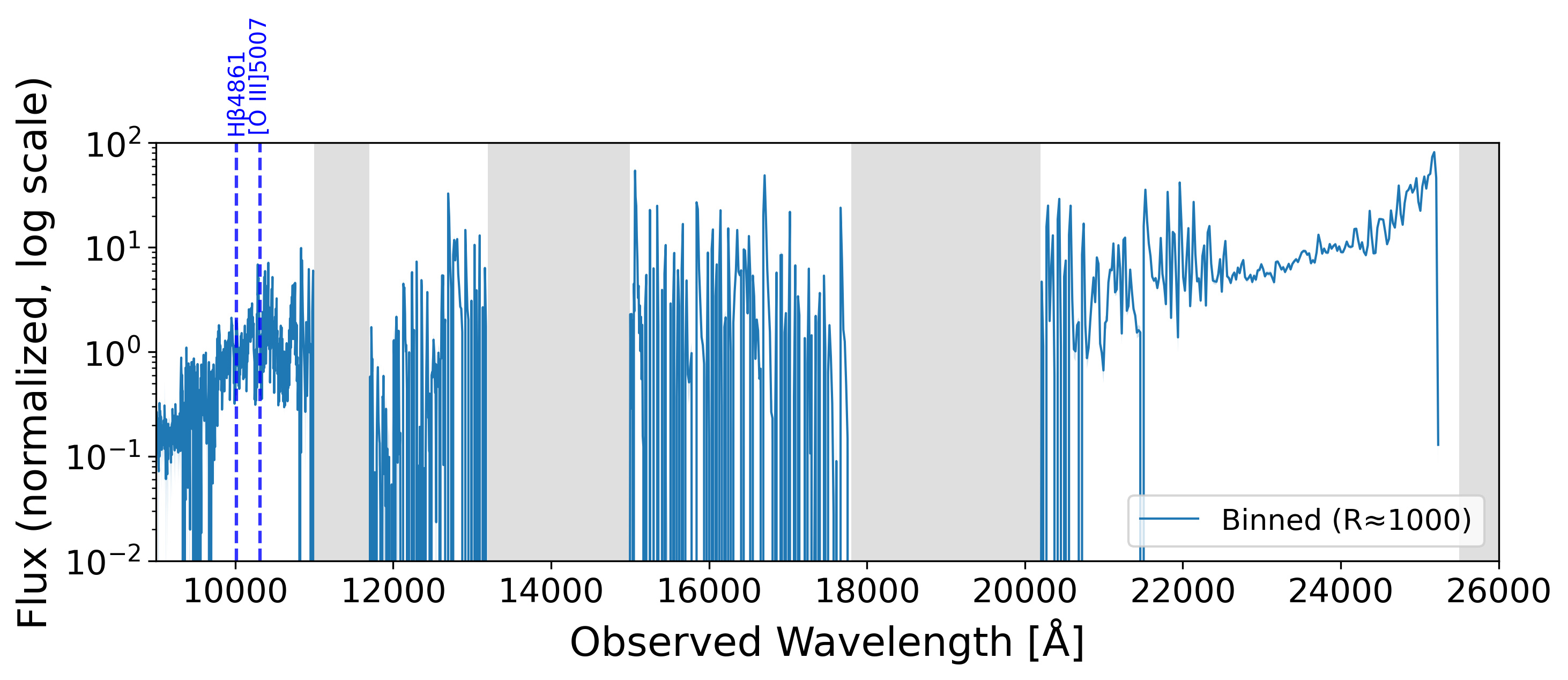}
  \vspace{1em}
  \parbox{\linewidth}{\small
    \vspace{1em}
\textbf{Extended Data Figure \arabic{extfig} | GNIRS 1D spectrum of the lens galaxy.}
The extracted orders were inverse-variance combined and rebinned to a constant resolving power of $R\simeq1000$; the flux is normalized by the median and plotted on a logarithmic $y$-axis. 
The $x$-axis shows observed wavelength in \AA\ (covering 0.9--2.6\,$\mu$m). 
Gray bands mark regions of strong telluric absorption (1.10--1.17, 1.32--1.50, 1.78--2.02, and $>$2.55\,$\mu$m). 
Blue dashed lines indicate the expected observed wavelengths of H$\beta$ and [O\,\textsc{iii}]~$\lambda5007$ at the photometric redshift $z_{\rm phot}=1.06$, but neither line is detected. 
[O\,\textsc{ii}]~$\lambda3727$ (0.768\,$\mu$m at this redshift) lies outside the effective GNIRS blue-camera range (0.9--2.5\,$\mu$m), and H$\alpha$ falls within a strong telluric band, so both remain unconstrained.}
  \label{fig:gnirs_spec}
\end{extfigure}

\begin{extfigure}[t]
  \centering
 \includegraphics[width=\textwidth]{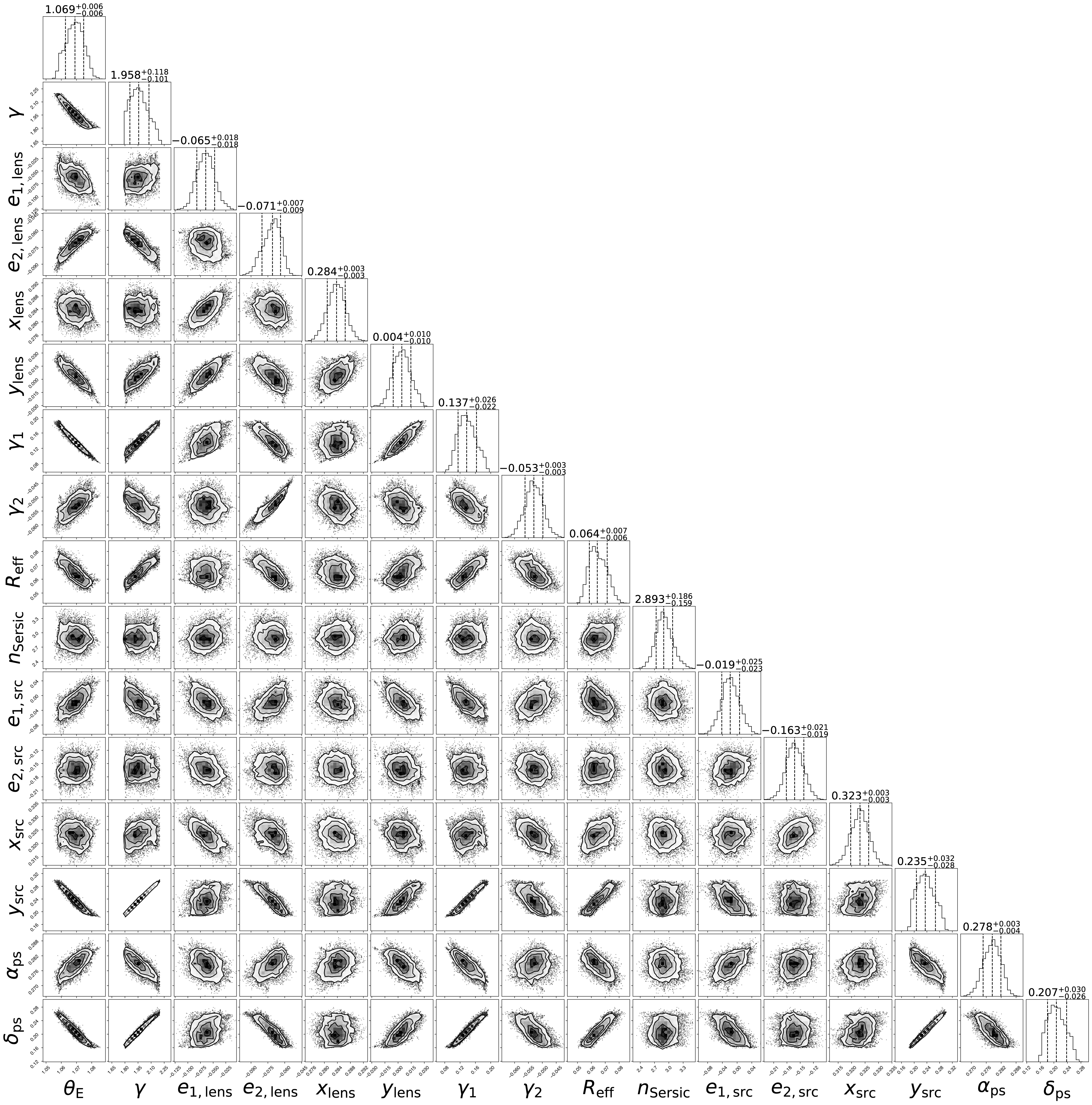}
  \vspace{1em}
  \parbox{\textwidth}{\small
    \vspace{1em}
\textbf{Extended Data Figure \arabic{extfig} | Posterior distributions from the MCMC lens modeling.}
Corner plot of the posterior distributions from the MCMC optimization of the lens and source parameters
for the \emph{Point Source + S\'ersic} model. 
The diagonal panels show the marginalized one-dimensional posterior distributions (with medians and 68\% credible intervals), 
while the off-diagonal panels display the covariances between pairs of parameters 
(e.g., lens mass normalization and ellipticity, external shear, source effective radius, S\'ersic index, and point-source flux). 
The posterior distributions are tightly constrained and reveal the expected parameter degeneracies in strong-lensing models.

  }
  \label{fig:lens_corner}
\end{extfigure}

\begin{extfigure}
  \centering
  \includegraphics[width=\textwidth]{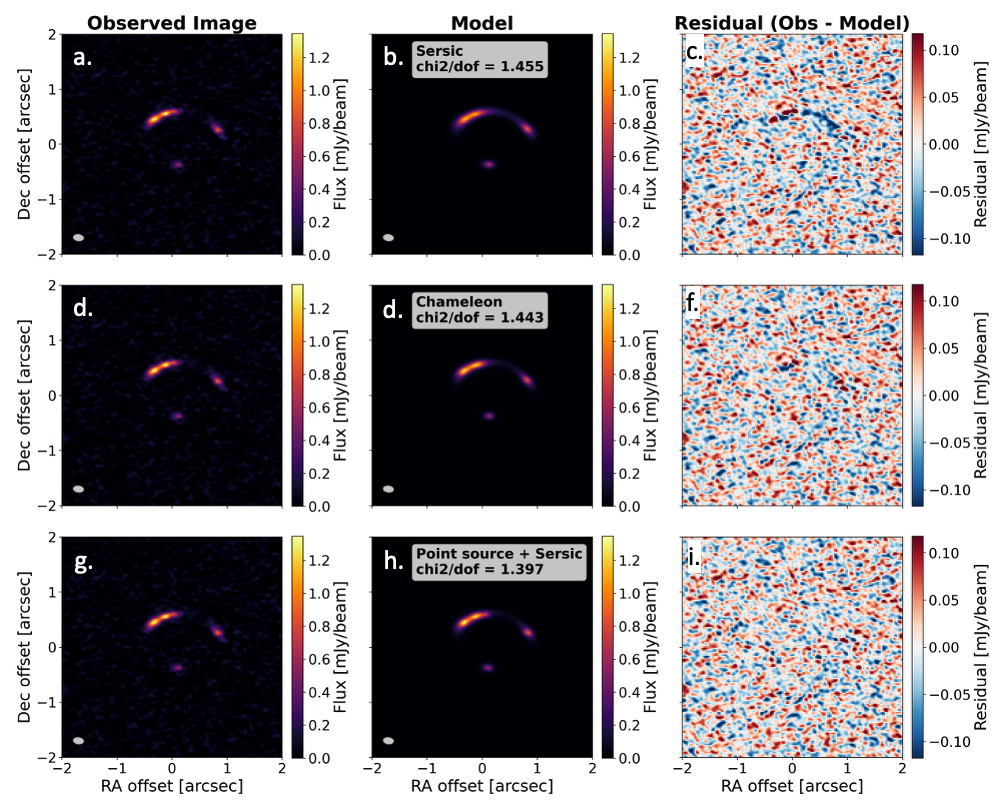}
  \vspace{1em}
  \parbox{\textwidth}{\small
    \vspace{1em}
\textbf{Extended Data Figure \arabic{extfig} | Comparison of source models in gravitational lens modeling.}
\textbf{a--c,} S\'ersic profile model: (a) observed ALMA Band~5 image, (b) best-fit model image ($\chi^2$/dof = 1.455), and (c) residual map (observed $-$ model). 
\textbf{d--f,} Chameleon profile model: (d) observed image, (e) model image ($\chi^2$/dof = 1.443), and (f) residual map. 
\textbf{g--i,} Point Source + S\'ersic model: (g) observed image, (h) model image ($\chi^2$/dof = 1.397), and (i) residual map. 
Color bars indicate flux density (mJy\,beam$^{-1}$) for the image and model panels, and residual intensity (mJy\,beam$^{-1}$) for the residual panels. 
The synthesized beam is shown as a gray ellipse in the lower-left corner of the observed and model panels. 
The Point Source + S\'ersic model achieves the best fit and yields residuals that are consistent with the image noise, significantly improving upon the single-profile models.
  }
  \label{fig:model_comparison}
\end{extfigure}

\begin{extfigure}[t]
  \centering
  \includegraphics[width=\textwidth]{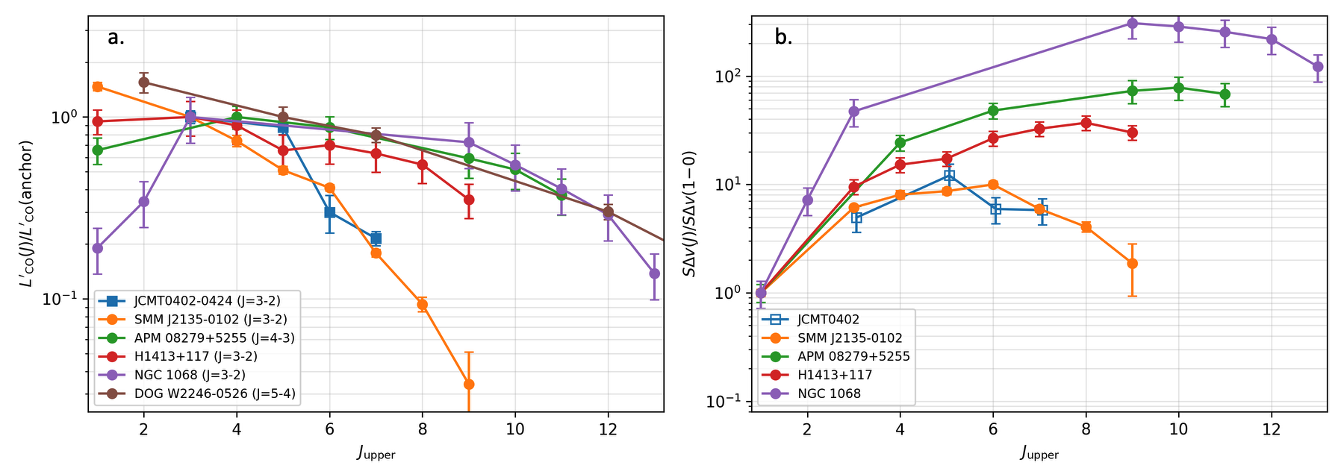}
  \vspace{1em}
  \parbox{\textwidth}{\small
    \vspace{1em}
\textbf{Extended Data Figure \arabic{extfig} | CO spectral line energy distributions (SLEDs).}
\textbf{(a)}  Within--object luminosity ratios $L'_{\rm CO}(J)/L'_{\rm CO}(\mathrm{anchor})$, where the anchor transition is chosen as CO(3--2) when available, otherwise the lowest available transition with $3\!\le\!J_{\rm upper}\!\le\!6$ (the chosen anchor is indicated in the legend; for the DOG W2246--0526 this is CO(5--4)). 
When only integrated fluxes $S\Delta v$ are reported, ratios are computed assuming $L' \propto (S\Delta v)/\nu^2$. 
\textbf{(b)} Velocity--integrated intensity ratios $S\Delta v(J)/S\Delta v(1\!-\!0)$; for JCMT0402--0424, CO(1--0) is inferred from CO(3--2) by adopting a typical DSFG range $r_{31}=L'_{\rm CO(3-2)}/L'_{\rm CO(1-0)}=0.4$--0.7, with the line showing the median value ($r_{31}=0.5$) and the error bars spanning this range; the luminous DOG W2246--0526 currently lacks a CO(1--0) measurement and is therefore not shown in this panel. 
Error bars indicate $1\sigma$ measurement uncertainties (or the adopted $r_{31}$ range for JCMT0402--0424).
The JCMT0402--0424 SLED rises to $J_{\rm upper}\!\sim$5 and declines toward higher $J$, closely resembling the compact--starburst template SMM~J2135--0102, whereas quasars (APM~08279+5255, H1413+117) and the luminous DOG W2246--0526 retain comparatively stronger mid--/high--$J$ excitation; NGC~1068 (compact nucleus) is shown as a nearby AGN template.
  }
  \label{fig:co_sled}
\end{extfigure}

\begin{extfigure}[t]
  \centering
  \includegraphics[width=\textwidth]{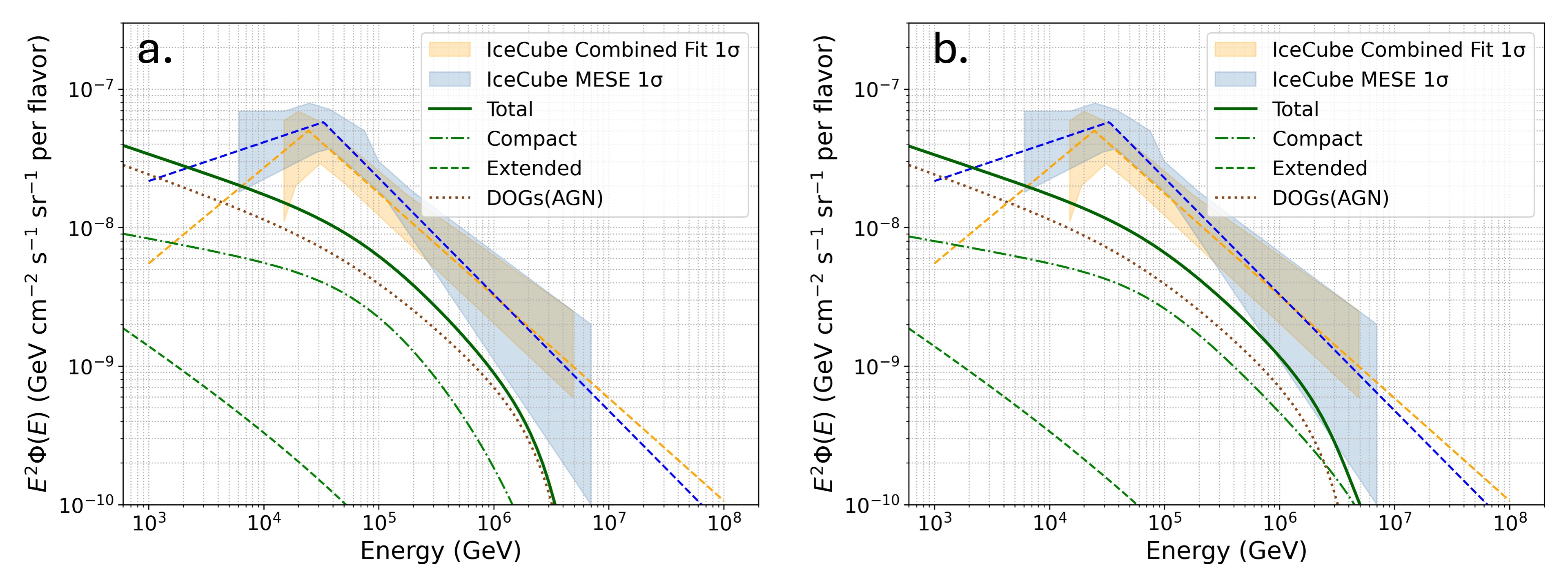}
  \vspace{1em}
  \parbox{\textwidth}{\small
    \vspace{1em}
\textbf{Extended Data Figure \arabic{extfig} | Diffuse neutrino flux from DSFGs\,+\,DOGs compared to IceCube: effect of the compact-core spectral form.}
Per-flavor $E^{2}\Phi(E)$ in the observed frame (all-flavor is $\times3$). 
Dark-green solid: total; green dash--dotted: compact DSFG core; green dashed: extended\,+\,LIRG; brown dotted: DOGs (AGN).
IceCube CF and MESE broken--power-law fits with $1\sigma$ bands are shown in orange and blue \citep{Abbasi2025}.
\textbf{(a)} Compact core as a single power law with a high-energy cutoff (CPL). 
\textbf{(b)} Compact core as a smoothly broken power law (SBPL). 
LFs, redshift evolution, efficiencies, and $f_c(z)$ are identical in both panels. 
Compared to the CPL in panel~(a), the SBPL in panel~(b) yields a more gradual steepening of the compact-core DSFG component and sustains the total spectrum to higher energies, while keeping a similar overall normalization.}
  \label{fig:dsfg_plus_agn}
\end{extfigure}

\clearpage


\end{document}